# Insight into the partitioning and clustering mechanism of rare-earth cations in alkali aluminoborosilicate glasses

Hrishikesh Kamat[1], Fu Wang[1, 2], Kristian Barnsley[3], John V. Hanna[3], Alexei M. Tyryshkin[4, *], Ashutosh Goel[1, *]

[1] Department of Materials Science and Engineering, Rutgers, The State University of New Jersey, Piscataway, New Jersey, USA, 08854

[2] School of Materials Science and Engineering, Southwest University of Science and Technology, Mianyang, 621010, PR China

[3] Department of Physics, University of Warwick, Coventry CV4 7AL, United Kingdom

[4] Department of Marine and Coastal Sciences, Rutgers, The State University of New Jersey, Piscataway, New Jersey, USA, 08854

* Corresponding authors:
Ashutosh Goel: email: ag1179@soe.rutgers.edu; Ph: +1-848-445-4512
Alexei M. Tyryshkin: email: tyrysh@gmail.com

**Abstract**

Rare-earth (RE) containing alkali aluminoborosilicate glasses find increasingly broad technological applications, with their further development only impeded by yet-poor understanding of coordination environment and structural role of RE ions in glasses. In this work we combine free induction decay (FID)-detected electron paramagnetic resonance (EPR), electron spin echo envelope modulation (ESEEM), and MAS NMR spectroscopies, to examine the coordination environment and the clustering tendencies of $RE^{3+}$ in a series of peralkaline aluminoborosilicate glasses co-doped with $Nd_2O_3$ (0.001–0.1 mol%) and 5 mol% $La_2O_3$. Quantitative EPR spectral analysis reveals three different $Nd^{3+}$ forms coexisting in the glasses: isolated $Nd^{3+}$ centers, dipole-coupled Nd clusters (Nd–O–X–O–Nd, where X = Si/B/Al), and spin-exchange-coupled Nd clusters, (Nd–O–Nd) and (Nd–O–La–O–Nd). Extensive RE clustering is observed at high $RE_2O_3$ concentrations, with more than 90% REs converting to dipole- and exchange-coupled Nd clusters already at $[RE_2O_3]$ = 0.01 mol%. ESEEM analysis of the EPR-detectable Nd centers indicates a Na/Si-rich environment (four $Na^+$ per $Nd^{3+}$) for the isolated $Nd^{3+}$ centers and the Na/Si/B-rich environment (2–3 $Na^+$ and 1–2 boron per each $Nd^{3+}$) for the dipole-coupled Nd clusters, while the EPR-undetectable exchanged-coupled RE clusters are predicted to exist in a Na/B-rich environment. The RE clustering induces nano-scale glass phase separation, while the Na/B-rich environment of the RE clusters implies a depletion of the same elements from the remaining host glass. Based on our results, we develop a mechanistic model that explains the high tendency of $RE^{3+}$ to form clusters in alkali aluminoborosilicate glasses.

1. **Introduction**

Our understanding of the mechanism of rare-earth (RE) clustering in $SiO_2$ glass and its suppression by $Al_2O_3$ and $P_2O_5$[1-5] has resulted in the development of RE-doped fiber amplifiers and lasers, thus, bringing a paradigm shift in the field of telecommunication. However, this has not been the case with multicomponent silicate glasses, where the technological development in the field of RE-containing glasses has been primarily confined to the academic realms. A major reason for the slow transition from academia to real-world applications is our poor understanding of the structural role of RE cations in these glasses (as discussed in more detail in the following paragraphs), thus, resulting in an ambiguity in their structure-property relationships. For example, $RE^{3+}$ are usually considered to play the role of network modifiers in the structure of alkali aluminoborosilicate glasses.[6, 7] Based on our conventional wisdom, network modifying cations are expected to depolymerize the silicate/aluminoborosilicate glass network by creating non-bridging oxygens (NBOs), and some evidence of it has indeed been reported.[6, 7] Such RE-induced glass depolymerization is expected to decrease the glass transition temperature ($T_g$) of the glasses. Nevertheless, $RE_2O_3$ additions to silicates, aluminosilicates, and aluminoborosilicate glasses have been shown to have the opposite effect of shifting their $T_g$ towards the higher side.[7-13] While the literature attributes the increase in $T_g$ to the stronger RE–O⁻ bonds being formed in the glass,[7, 8, 11] this is simply a hypothesis as no structural evidence has ever been reported to this effect.

The lack of understanding about the fundamental science governing the role of $RE^{3+}$ in the structure of multicomponent silicate glasses becomes a bottleneck in explaining the trends observed in functional glasses. For example, $MoO_3$-loading in an alkali aluminoborosilicate-based nuclear waste glass has been shown to be highly dependent on the nature and concentration of RE present in the glass.[14] However, the reason for this dependence is debatable as it is not explicit if RE$^{3+}$ simply act as network modifiers in these glasses or if they induce clustering and phase separation.[15, 16] Other examples, where our lack of understanding about the role of rare-earth cations in the structure of multicomponent silicate glasses has slowed down their further technological development, can be quoted from the field of biomaterials[17] and solid-state lasers or optical amplifiers.[18] Thus, the present study is directed towards understanding the structural role of rare-earth cations in alkali aluminoborosilicate glasses. The overarching goal is to find answers to the following questions: Do RE$^{3+}$ distribute homogeneously or heterogeneously in the matrix of alkali aluminoborosilicate glasses? In either case, what is their chemical coordination environment in multicomponent glasses, and what mechanisms define their distribution in the glass network?

Accordingly, we have selected the $Na_2O$-$Al_2O_3$-$B_2O_3$-$SiO_2$ based quaternary glass system with Na/Al > 1. The choice of glass system has been made considering that it is the backbone of the specialty glass industry and a potential candidate for several upcoming technological innovations. As an example, cover glasses used as the outer contact surface of touch-screen electronic displays are designed primarily in the $Na_2O$-$(K_2O)$-$Al_2O_3$-$B_2O_3$-$(P_2O_5)$-$SiO_2$ systems where alkali/$Al_2O_3 \geq$ 1.[19] Similarly, borosilicate-based nuclear waste glasses are generally designed in $Na_2O$-$Al_2O_3$-$B_2O_3$-$SiO_2$ systems with Na/Al > 1.[15, 20, 21] Further, RE$^{3+}$-doped alkali aluminoborosilicate glasses have been proposed as broad emission bandwidth laser media suitable for use at powers higher than petawatt or exawatt levels.[22] Therefore, insights into the structural role of rare-earth cations in the $Na_2O$-$Al_2O_3$-$B_2O_3$-$SiO_2$ systems are not only of scientific interest but also have a tangible impact on the design of functional glasses for a broad spectrum of technological applications.

**2. Distribution/partitioning of rare-earth cations in alkali aluminoborosilicate glasses – Literature review**

Two opposing viewpoints are prevailing in the glass science community. The first viewpoint[23-26] suggests that $RE^{3+}$ induce nano-scale phase separation in homogenous peralkaline (Na/Al > 1) aluminoborosilicate glasses leading to the segregation of borate-rich and silicate-rich phases. $RE^{3+}$ are believed to preferentially dissolve in the borate-rich region forming RE-metaborate-like structures ($1BO_4$:RE:$2BO_3$). Once the $RE^{3+}$ concentration in the glass exceeds the limit of saturation in the borate-rich region (i.e., $[RE_2O_3] = 1/3[B_2O_3]$, according to reference[23]), they enter the silicate-rich region forming RE-clusters. On the other hand, the second viewpoint argues against the clustering tendency of $RE^{3+}$ or the $RE^{3+}$ induced nano-scale phase separation in the glass network. This hypothesis suggests that $RE^{3+}$ enter the depolymerized region of a homogenous peralkaline aluminoborosilicate glass and establish their local environment by directly coordinating with the alkali/alkaline earth ions for charge compensation.[6, 27]

Both viewpoints are based on thorough investigations involving a suite of spectroscopic and microscopic techniques like electron energy loss spectroscopy (EELS), extended X-ray absorption fine structures (EXAFS), fluorescence spectroscopy, nuclear magnetic resonance (NMR), Raman spectroscopy, and transmission electron microscopy (TEM).[6, 23-27] However, neither of these techniques provides an atomic-scale picture of RE clustering and their coordination environment (beyond the first neighbor shell, typically oxygen) in alkali aluminoborosilicate glasses. Also, all these techniques lack in their ability to selectively probe the $RE^{3+}$ centers, quantify possible RE speciation (isolated centers versus RE clusters) and estimate the extent of RE clustering in glasses. Therefore, to gain deeper insight into the local environment and distribution of $RE^{3+}$ over a length scale longer than few Å (beyond next neighbor distances), we need a spectroscopic technique that is selective in its ability to probe selective RE ions, has an improved concentration sensitivity, and a higher spatial resolution.

3. **Electron paramagnetic resonance (EPR) is a method of choice to study $RE^{3+}$ environment in glasses**

EPR spectroscopy is a technique capable of probing paramagnetic ions in solids. It offers several important advantages over other techniques (e.g., EXAFS, fluorescence spectroscopy, NMR, Raman spectroscopy) conventionally used for characterization of $RE^{3+}$ speciation and clustering in glasses.[1, 2, 24, 28-30]

(1) EPR has a superior concentration sensitivity, benefiting from its ability to selectively probe the paramagnetic $RE^{3+}$ centers only, without an overwhelming background contribution from other (often more abundant) centers and defects in glasses. The EPR sensitivity is better than 1 ppb (part per billion), or about 3–4 orders of magnitude higher than in other techniques, like UV-Vis, Raman, NMR, and EXAFS. EPR can be used to probe broad ranges of RE concentrations from 0.0001 to 1 mol%.

(2) Unlike other techniques, EPR allows an accurate evaluation of the number of spins contributing to the signal (known as spin counting) without any underlying assumptions.

(3) Pulsed EPR techniques (as used in this work) offer superior spectral resolution which allows resolving multiple $RE^{3+}$ species (e.g., isolated $RE^{3+}$ centers versus RE clusters) if present in the glass.[31]

(4) Electron spin echo envelope modulation (ESEEM) is one of many pulsed EPR methods that can probe the coordination environment of $RE^{3+}$ ions by measuring electron-nuclear hyperfine interactions with magnetic nuclei in the glass matrix.[32] ESEEM can often probe the second and third coordination shells around $RE^{3+}$, which is far beyond of what is accessible when using EXAFS and optical absorption spectroscopy.

Given these advantages, EPR has been previously employed for investigating $RE^{3+}$ speciation and coordination environment in silicate and phosphate-based glasses.[3-5, 33-37] Specifically, in silica-rich glasses, several articles published by Sen and co-workers,[3, 5, 37] Saitoh et al.,[4] and Arai et al.[33] have reported on the spatial distribution (RE–RE distances) of the dopant $RE^{3+}$ and their next-nearest neighbor atoms in glasses. Here, we build up on these previous works to examine the $RE^{3+}$ speciation in alkali aluminoborosilicate glasses (Na/Al > 1).

In this study, we combine pulsed EPR spectroscopy (Free induction decay (FID)-detected EPR and ESEEM) and $^{11}B/^{27}Al$ magic angle spinning nuclear magnetic resonance (MAS NMR) spectroscopy to probe the speciation and local environment of rare-earth cations ($Nd^{3+}$) in peralkaline (Na/Al > 1) aluminoborosilicate glasses. From quantitative spectral EPR analysis we determine that $Nd^{3+}$ is present in three different forms in our glasses: isolated $Nd^{3+}$ centers, dipole-coupled Nd clusters, and spin-exchange-coupled Nd clusters. The fractional concentrations of each species are determined as a function of total Nd concentration in the glass and also after adding additional 5 mol% of diamagnetic $La^{3+}$. Extensive RE clustering (more than 90% of total RE) is observed at RE concentrations as low as 0.01 mol% in the glasses. From ESEEM spectroscopy and supported by $^{11}B$ and $^{27}Al$ MAS NMR, we find that isolated $Nd^{3+}$ centers are preferentially formed in the Na/Si-rich coordination, while the Nd clusters are formed in the Na/B-rich coordination. Based on these results, we propose the structural models for isolated and clustered RE species in the peralkaline aluminoborosilicate glasses and develop a charge compensation model to explain the high tendency of $RE^{3+}$ to form clusters.

## 4. Experimental details

### 4.1 Rationale for glass composition design

A peralkaline (Na/Al > 1) sodium aluminoborosilicate glass with a batched composition of 25 $Na_2O$– 10 $Al_2O_3$– 10 $B_2O_3$– 55 $SiO_2$ (mol%) has been selected for $RE_2O_3$ addition. This glass composition is chosen based on the criterion proposed by Li et al.,[23] specifically $[Na_2O]_{ex}/[B_2O_3] > 0.5$, where $[Na_2O]_{ex} = [Na_2O] - [Al_2O_3]$, for designing homogenous peralkaline aluminoborosilicate glasses with no detectable (microscopic) phase separation. The spectroscopic examination by Wu and Stebbins[38] have confirmed the validity of this criterion, demonstrating that peralkaline sodium aluminoborosilicate glasses with $[Na_2O_{ex}]/[B_2O_3] = 1.7$ (similar to the one reported in the present study $[Na_2O_{ex}]/[B_2O_3] = 1.5$) are indeed homogeneous.

$Nd^{3+}$ are chosen as EPR spin probes for investigating the $RE^{3+}$ speciation and their local environments in alkali aluminoborosilicate glasses. Accordingly, the baseline glass (mol%, 25 $Na_2O$– 10 $Al_2O_3$– 10 $B_2O_3$– 55 $SiO_2$) has been doped with varying concentrations of $Nd_2O_3$ in the range from 0.001 to 0.1 mol% to produce a series of glasses labeled as Ndx, where 'x' is the $Nd_2O_3$ concentration in mol% (Table 1).

Our preliminary experiments on Nd-containing glasses reveal that the $Nd_2O_3$ concentrations greater than 0.1 mol% are unsuitable for pulsed EPR experiments, resulting in: (1) significant EPR line broadening, due to excessive dipole-dipole interactions between paramagnetic $Nd^{3+}$ ions, making it impossible to resolve the individual contributions from different Nd species present in the glasses; and (2) too short $T_2$ spin relaxation times of $Nd^{3+}$ (e.g., $T_2 < 100$ ns, below the detection limit of our EPR experiment), owing again to strong Nd–Nd dipole-dipole interactions and also strong spin-exchange interactions, resulting in most of Nd centers to become EPR-invisible (undetectable). To overcome this experimental limitation and to examine RE speciation at concentrations higher than 0.1 mol%, a second series of glasses has been synthesized where $Nd_2O_3$ (x = 0.001–0.1 mol%) has been co-doped with $La_2O_3$ [(5 − x) mol%]. Accordingly, the second glass series has a composition: 23.75 $Na_2O$– 9.50 $Al_2O_3$– 9.50 $B_2O_3$– 52.25 $SiO_2$– (5.00 − x) $La_2O_3$– x $Nd_2O_3$ (mol%) and labeled as LaNdx. $La^{3+}$ is used as it is commonly viewed as a diamagnetic substitute of $Nd^{3+}$ in glass chemistry due to their similar coordination environments (6–8 oxygens surrounding the RE ions) and their similar cation field strengths.[39,40] Table 1 presents the batched compositions of all glasses examined in this work.

Table 1: Batched composition (in mol%) of melt-quenched glasses examined in this work. Relative fractions of trigonal $BO_3$ ($N_3$ = [$BO_3$]/([$BO_3$] + [$BO_4$])) and tetrahedral $BO_4$ ($N_4$ = [$BO_4$]/([$BO_3$] + [$BO_4$])) boron units estimated from $^{11}B$ MAS NMR spectra of $Nd_2O_3$-free base glasses, Nd0 and LaNd0 (Figure 5b).

| Glass ID | Na$_2$O | Al$_2$O$_3$ | B$_2$O$_3$ | SiO$_2$ | Nd$_2$O$_3$ | La$_2$O$_3$ | N$_3$ | N$_4$ |
|---|---|---|---|---|---|---|---|---|
| La$_2$O$_3$-free | | | | | | | | |
| Nd0 | 25.00 | 10.00 | 10.00 | 55.00 | - | - | 0.34 | 0.66 |
| Nd0.001 | 25.00 | 10.00 | 10.00 | 55.00 | 0.001 | - | | |
| Nd0.01 | 25.00 | 10.00 | 10.00 | 54.99 | 0.01 | - | | |
| Nd0.05 | 24.99 | 10.00 | 10.00 | 54.96 | 0.05 | - | | |
| Nd0.1 | 24.98 | 10.00 | 10.00 | 54.92 | 0.1 | - | | |
| La$_2$O$_3$-containing | | | | | | | | |
| LaNd0 | 23.75 | 9.5 | 9.5 | 52.25 | - | 5.00 | 0.63 | 0.37 |
| LaNd0.001 | 23.75 | 9.5 | 9.5 | 52.25 | 0.001 | 4.999 | | |
| LaNd0.01 | 23.75 | 9.5 | 9.5 | 52.25 | 0.01 | 4.99 | | |
| LaNd0.05 | 23.75 | 9.5 | 9.5 | 52.25 | 0.05 | 4.95 | | |
| LaNd0.1 | 23.75 | 9.5 | 9.5 | 52.25 | 0.1 | 4.9 | | |

### 4.2 Glass synthesis by melt-quenching

High purity powders of $Na_2SiO_3$ (Alfa Aesar > 99%), $Al_2O_3$ (Sigma-Aldrich ≥ 98%), $H_3BO_3$ (Alfa Aesar ≥ 98%), $SiO_2$ (Alfa Aesar ≥ 99.5%), $La_2O_3$ (Sigma-Aldrich, 99.9%), and $Nd_2O_3$ (Alfa

Aesar, 99%) were used as precursors for glass synthesis. The batches corresponding to 50 g of glass were melted in 90% Pt – 10% Rh (wt.%) crucibles covered with a Pt lid (to reduce losses due to volatilization). The $La_2O_3$-free batches were melted at 1773 K while $La_2O_3$-containing at 1823 K for 1 hour. The melts were quenched on a copper plate followed by annealing for 2 hours at temperatures set about 50 K lower than their respective glass transition temperatures ($T_g$ = 824 K for $La_2O_3$-free and 840 K for $La_2O_3$-containing baseline glasses, as determined by differential scanning calorimetric curves at a heating rate of 20 K/min — STA449F5 Jupiter, NETZSCH, Figure S1 — Supplementary Information). The post-annealed glass samples were visibly transparent and amorphous when examined by X-ray diffraction (not shown).

### 4.3 *Electron paramagnetic resonance (EPR) spectroscopy*

Pulsed EPR experiments were performed on the monolith glass samples diced to fit into standard 4 mm EPR tubes. A Bruker EPR spectrometer (Elexsys580E) operating at X-band frequency (9.7 GHz) was used with a dielectric resonator (EN-4118X-MD4) and a helium-flow cryostat (Oxford CF935). All pulsed EPR experiments were performed at 4.6 K.

Field-sweep EPR spectra were recorded using an FID-detected EPR experiment by applying one excitation π/2 pulse of duration 400 ns to 1 μs and integrating an FID signal over 400–800 ns window starting at a time point 64 ns after the excitation pulse (e.g., the dead-time in our EPR experiments). FID-detected EPR has many advantages against the often-used echo-detected EPR by providing improved spectral sensitivity and, more importantly, by eliminating any distortions to EPR spectral lineshape originating from interfering ESEEM effects.[41-43] The echo-detected EPR spectra measured for two of our samples are compared with their FID-detected counterparts in Figure S2, illustrating the spectral lineshape distortions and the significant loss in signal intensity in the echo-detected EPR spectra as compared to the FID-detected EPR spectra. To allow a quantitative comparison of the EPR signal intensities and lineshapes measured for different samples, the as-recorded FID-detected EPR spectra were normalized per sample weight and further corrected for differences in their measured $T_2$ spin relaxation times while taking into account the initial dead-time (64 ns) and the FID detected window (400–800 ns) in each experiment.

ESEEM experiments were performed using a two-pulse (Hahn) echo pulse sequence ($\pi/2-\tau-\pi-\tau-$echo), with $\pi/2$ and $\pi$ pulses set to 16 ns and 32 ns, respectively. Before Fourier transformation (FT), the measured ESEEM time-domains were normalized by dividing the experimental data with the fit exponential curves and subtracting a unit baseline. After FT, the ESEEM spectra were phase-corrected to compensate for the missing dead time (the initial $\tau = 104$ ns in all ESEEM experiments) and finally plotted in the cosine FT mode. All FID-detected EPR and ESEEM spectral simulations were performed using the EasySpin[44] toolbox developed for MATLAB. The same normalization procedure was applied to the simulated ESEEM time domains to allow their direct comparison with the experimental FT ESEEM spectra.

### 4.4 $^{11}B$ and $^{27}Al$ MAS NMR spectroscopy

$^{11}B$ and $^{27}Al$ MAS NMR spectra were acquired on the baseline $La_2O_3$-free (Nd0) and $La_2O_3$-containing glass (LaNd0) to examine the impact of adding $La_2O_3$ on boron and aluminum speciation in the glass structures. $^{11}B$ MAS NMR spectra were collected at 14.1 T and 11.7 T using Bruker Avance-II+-600 and Avance-III-500 MHz spectrometers. Measurements were performed using a Varian 4 mm MAS NMR probe at a spinning frequency of 12 kHz. The data were acquired using a single pulse method, and both pulse time calibration and chemical shift referencing were achieved using solid $NaBH_4$. All $^{11}B$ isotropic chemical shifts were referenced against the IUPAC standard $BF_3Et_2O$ ($\delta_{iso} = 0.0$ ppm) using $NaBH_4$ as a secondary reference ($\delta_{iso} = -42.06$ ppm).

$^{27}$Al MAS NMR spectra were collected at 14.1 T and 9.4 T using Bruker Avance-II+-600 and Bruker HD-400 MHz spectrometers. The FID curves were measured via the single-pulse method using a Bruker 3.2 mm MAS NMR probe, enabling a sample spinning frequency of 20 kHz. Both pulse time calibration and $^{27}$Al isotropic chemical shift referencing were achieved using 1.1 M Al(NO$_3$)$_3$, the primary IUPAC standard ($\delta_{iso}$ = 0.0 ppm). The variable field $^{11}$B and $^{27}$Al MAS NMR data were simulated using the QuadFit simulation program,[45] where the multi-field approach was used to further constrain the simulation parameters.

## 5. Results and Discussion

### 5.1 FID-detected EPR

The FID-detected EPR spectra of all La$_2$O$_3$-free (Ndx) and La$_2$O$_3$-containing (LaNdx) glasses examined in the present study are shown in Figure 1a and Figure 1b, respectively. The spectra reveal complex transformations, with both EPR spectral intensities and spectral lineshapes changing while increasing the Nd$_2$O$_3$ concentrations from 0.001 to 0.1 mol% (Figure 1a), and also after the addition of 5 mol% La$_2$O$_3$ (Figure 1b). A detailed analysis of these spectral transformations, as presented next, reveals that Nd$^{3+}$ ions co-exist in several structural forms in these glasses, including isolated Nd$^{3+}$ centers and two types of Nd clusters. The isolated Nd$^{3+}$ centers are observed to be the dominant species only in the glasses with the lowest RE$_2$O$_3$ concentrations (0.001 mol%), and the Nd cluster population grows rapidly to become dominant at RE$_2$O$_3$ concentrations higher than 0.01 mol%. These concentration-dependent changes in Nd speciation are a cause of the striking transformations in the EPR spectral lineshapes observed in Figure 1.

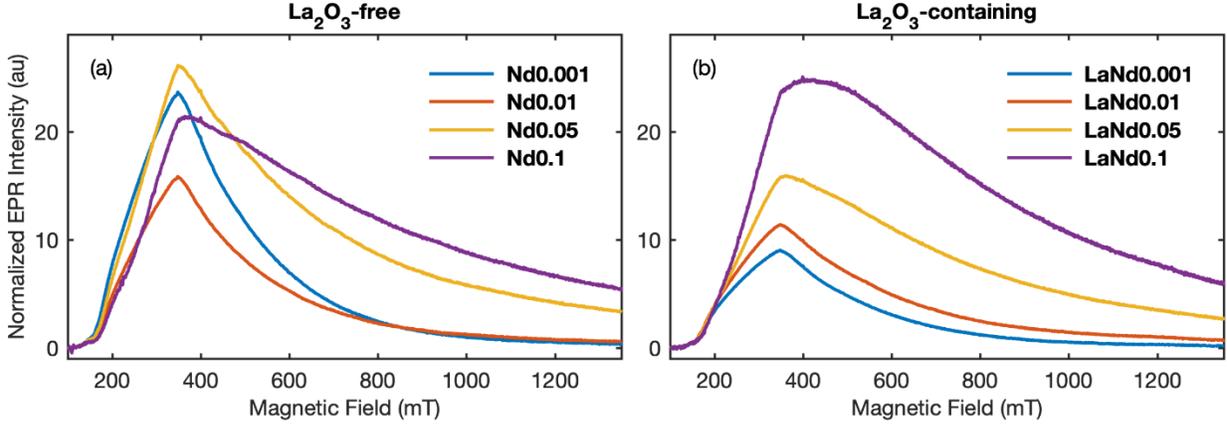

**Figure 1**: FID-detected EPR spectra of peralkaline sodium aluminoborosilicate glasses doped with varied concentrations of [$Nd_2O_3$] and with (a) none, and (b) 5 mol% of [$La_2O_3$], as labeled in the figure legend. The spectral intensities in each sample were normalized per unit sample weight and also corrected for differences in spin relaxation times and other instrumental settings as described in Experimental section. All spectra were measured at 4.6 K.

*5.1.1  Low $Nd_2O_3$ concentrations (0.001–0.01 mol%): Isolated $Nd^{3+}$ centers*

As seen in Figure 1, the EPR signals measured at low $Nd_2O_3$ concentrations (0.001–0.01 mol%), in both $La_2O_3$-free and $La_2O_3$-containing glasses, maintain the same spectral lineshape, and only the overall signal intensity changes between the samples. This EPR signal has been assigned to the isolated $Nd^{3+}$ centers that are expected to be in abundance at low $Nd_2O_3$ doping levels. The signal assignment to isolated $Nd^{3+}$ centers is further supported by the EPR simulation shown in Figure 2a (the dotted black line), performed assuming an effective spin S = 1/2 with a rhombic g-factor (3.5, 1.94, 1.0) and a significant g-factor strain, $\delta_g$ = (0.8, 0.3, 0.5). The derived g-factor values agree well with the values reported for isolated $Nd^{3+}$ centers in $Y_2SiO_5$ and La-nicotinate crystals doped with 0.001 at% $Nd_2O_3$, where $Nd^{3+}$ are known to occupy substitutional $C_1$-symmetry sites, being directly coordinated by six to eight oxygen atoms and with Si/C atoms in the next-nearest shell.[46-48] The derived g-factor values are characteristic of $Nd^{3+}$ ions with the ground-state Kramer doublet preferentially defined by spin projections ±1/2 in the $^4I_{9/2}$ multiplet.[49, 50] The broad g-factor strain distribution in our EPR simulation reflects a high disorder in coordination environment of isolated $Nd^{3+}$ ions in the glasses. Hereafter, this EPR signal from isolated $Nd^{3+}$ centers as observed in the glasses doped with low concentrations of $Nd_2O_3$ will be referred to as the signal S1.

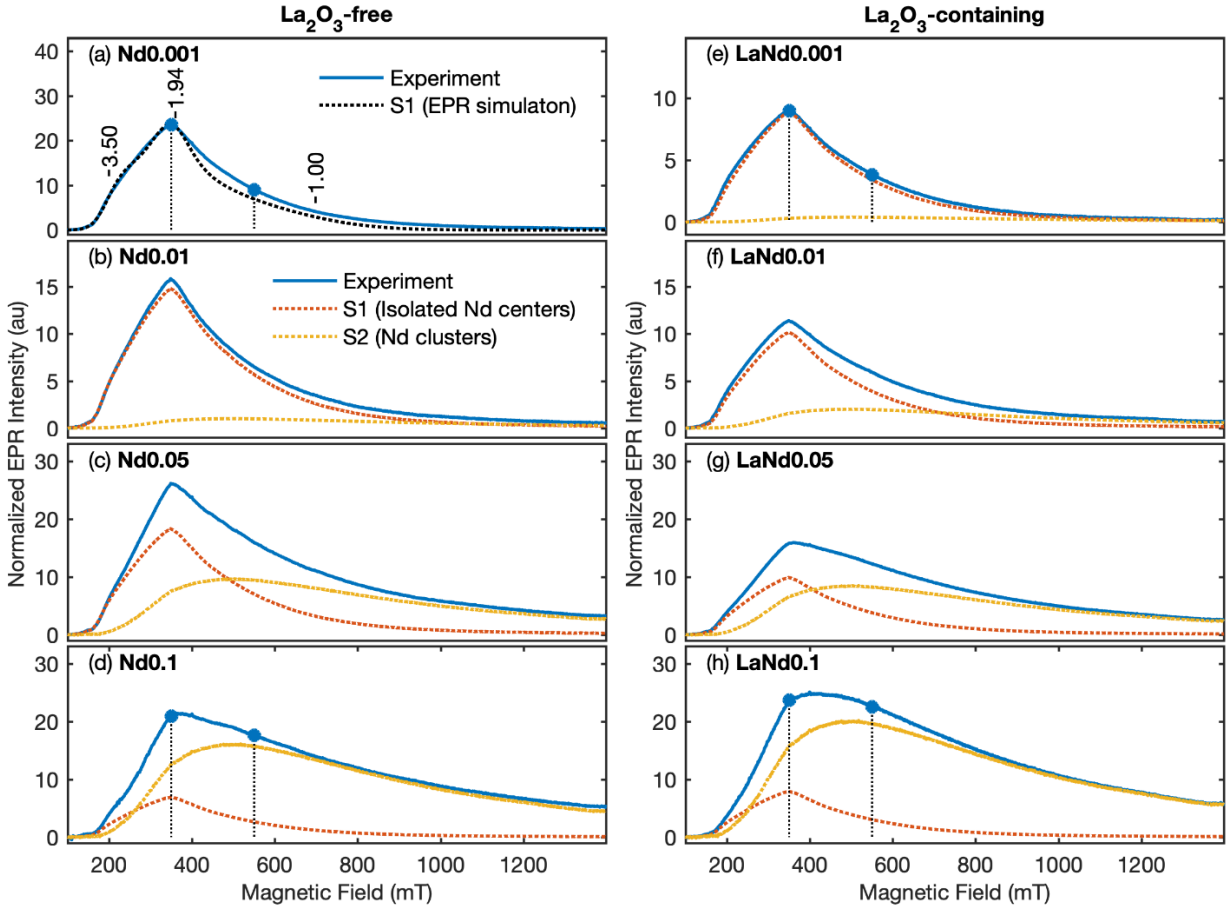

**Figure 2:** Deconvolution of the measured EPR spectra as weighted sums of two contributing EPR signals: the S1 signal (dash red) from isolated $Nd^{3+}$ centers, and the S2 signal (dash yellow) from dipole-coupled Nd–Nd clusters. The experimental spectrum in Nd0.001 (a), with the lowest $Nd_2O_3$ concentration and no added $La_2O_3$, was taken to be purely from isolated $Nd^{3+}$ centers (the pure S1 signal). The S2 signal was determined from the EPR spectrum measured for LaNd0.1 (h), after subtracting a small contribution of the S1 signal. The simulated EPR signal for isolated $Nd^{3+}$ centers shown in (a) assumes a rhombic electron g-factor (3.5, 1.94, 1.0). The blue circles on the spectral curves in (a, d, e, and h) indicate the magnetic field positions where the ESEEM experiments were performed.

Notably, the intensity of signal S1 does not scale proportionally with total $Nd_2O_3$ concentration in glasses (Figure 1). The scaling discrepancy is most obvious when comparing two $La_2O_3$-free glasses, Nd0.001 and Nd0.01, where a 10× increase in $Nd_2O_3$ concentration results in a 1.5× decrease in the S1 signal intensity (Figure 1a). Similarly, when comparing glasses, Nd0.001 and LaNd0.001, the addition of 5 mol% $La_2O_3$ is observed to cause a substantial (2.6×) decrease in the S1 signal intensity even though $Nd_2O_3$ concentration is kept constant (cp. Figures 1a and 1b). This non-linear dependence of the S1 signal intensity on $Nd_2O_3$ and $La_2O_3$ concentrations provides a strong evidence of $Nd^{3+}$ ions co-existing in multiple spectroscopically distinct forms in the investigated glasses. Some of these forms (e.g., isolated S1 centers) contribute fully to the EPR spectra, while other forms (see discussion below) remain EPR-invisible. The S1 signal shows up at its full strength only in the glass Nd0.001, i.e., with the smallest concentration of $RE_2O_3$. Here, all $Nd^{3+}$ centers are in the isolated S1 form and therefore, they fully contribute to the EPR signal intensity. The situation changes significantly when increasing the $Nd_2O_3$ concentration to 0.01 mol% (Nd0.01), or after adding 5 mol% $La_2O_3$ (LaNd0.001). In these two glasses only a small fraction of total Nd remains in the isolated S1 form which explains the disproportionally weak EPR signals, and a larger fraction of $Nd^{3+}$ is converted to EPR-invisible form(s), not contributing to the measured EPR spectra. More evidence of the complex speciation of $Nd^{3+}$ in the investigated glasses, involving several EPR-visible and EPR-invisible $Nd^{3+}$ forms, is found in the EPR spectra of glasses with $[Nd_2O_3] \geq 0.05$ mol%, as discussed next. The nature of EPR-invisible $Nd^{3+}$ forms is discussed in the section 5.1.6.

### 5.1.2 High $Nd_2O_3$ concentrations (0.05–0.1 mol%)

As $Nd_2O_3$ concentration increases further, the integrated EPR signal intensity keeps increasing but at a much slower rate than expected if all added $Nd^{3+}$ ions were EPR-visible. With $Nd_2O_3$ concentration changing by 100× from 0.001 mol% to 0.1 mol%, the integrated EPR signal intensity increases only by 2× in $La_2O_3$-free glasses and 5× in $La_2O_3$-containing glasses (Figure 1). The slower-than-expected growth of the EPR signal intensity in glasses doped with higher concentrations of $Nd_2O_3$ (0.01–0.1 mol%) further supports our conclusion about the complex speciation of $Nd^{3+}$, where only a small fraction of added $Nd^{3+}$ contributes to the observed EPR spectra, and a large fraction of $Nd^{3+}$ is converted into some EPR-invisible Nd form(s).

In addition to the changes in the signal intensity, the EPR spectral lineshape also changes at high $Nd_2O_3$ concentrations (Figure 1). In both $La_2O_3$-free and $La_2O_3$-containing glasses, the EPR lineshape gradually broadens out, and its center of gravity shifts to higher magnetic fields as $Nd_2O_3$ concentration increases above 0.01 mol%. Similar spectral transformations have been previously reported for $Nd^{3+}$ and other $RE^{3+}$ in silica- and non-silica-based glasses doped with elevated $RE_2O_3$ concentrations (0.01–1 mol%).[3, 5, 33-37, 51, 52] Two models have been proposed in the literature to explain these spectral transformations. One model emphasizes an intrinsic disorder of local RE environments in glasses and, thus, attributes the observed spectral transformations to a gradual change in the average atomic coordination environment of $RE^{3+}$ at high $RE_2O_3$ concentrations.[34-36, 52] The second model suggests the formation of $RE^{3+}$ clusters at high $RE_2O_3$ concentrations, e.g., through the formation of RE–O–RE like short-range linkages, where two or more $RE^{3+}$ are interconnected by bridging oxygens.[3, 5, 37, 51] Spin dipolar interactions between two or more $RE^{3+}$ ions within such short-range clusters (e.g., RE–O–RE, RE–O–X–O–RE) can then explain both the observed EPR lineshape broadening and the up-field center shift, as further confirmed in the detailed EPR spectral simulations.[37] Each of these models, or a combination of both, can as well be responsible for the EPR spectral transformations observed in our glasses at higher $Nd_2O_3$ concentrations (Figure 1). To help interpret these EPR spectral transformations, we performed the EPR spectral deconvolution analysis, as discussed in the next section.

*5.1.3 Two-component decomposition of the FID-detected EPR spectra*

In the previous EPR studies, the highly distorted EPR spectral lineshapes measured using echo-detected EPR had severely limited their quantitative spectral analysis.[3-5, 34-37, 51, 52] In the present study, the highly accurate lineshapes measured using FID-detected EPR allow us for a more in-depth spectral analysis, thus, providing further insight into $Nd^{3+}$ speciation in the investigated glasses.

A closer inspection of the spectral transformations in Figure 1 suggests that the EPR spectra in all eight glasses can be accurately described as simple weighted sums of two basic spectral components, hereafter referred to as the S1 and S2 signals, as demonstrated in Figure 2. The S1 signal originates from isolated $Nd^{3+}$ centers (see Section 5.1.1), and this signal is the only spectral component in the EPR spectrum of glass Nd0.001 (Figure 2a). The second signal, S2, is derived from the EPR spectrum of glass LaNd0.1 after subtracting a small contribution of the S1 signal (see the yellow trace for the derived S2 signal and the red trace for the subtracted S1 component in Figure 2h). The rationale behind this procedure for deriving the S2 component from the spectrum of LaNd0.1 is the observation that this spectrum demonstrates the most noticeable change from the S1 signal (e.g., the broadest EPR lineshape and the largest up-field shift) as compared to other samples. In other words, the spectrum of glass LaNd0.1 has the largest contribution from the signal S2 and a minimal contribution from the signal S1. The nature of the thus-derived S2 signal and its assignment to a specific Nd-related species is discussed in section 5.1.5.

The decomposition of the EPR spectra as weighted sums of the two spectral components, S1 (red) and S2 (yellow), are shown for all $La_2O_3$-free and $La_2O_3$-containing glasses in Figure 2b–2h. The S1 component (isolated $Nd^{3+}$ centers) is dominant only at low $Nd_2O_3$ concentrations (0.001–0.01 mol%), and the equilibrium shifts in favor of the S2 signal at $Nd_2O_3$ concentrations $\geq 0.05$ mol%. The addition of 5 mol% $La_2O_3$ induces further changes into the equilibrium of the two spectral components, as the S1 signal becomes further suppressed at the expense of the S2 signal.

*5.1.4 Three species model for $Nd^{3+}$ ions in glasses*

By integrating the signal intensities of the S1 and S2 components extracted from each EPR spectrum in Figure 2, we determine the absolute concentrations of S1 and S2 centers in each glass. The derived absolute S1 and S2 concentrations are plotted as a function of $Nd_2O_3$ and $La_2O_3$ concentrations in Figure 3a, and they are converted to the fractional Nd concentrations in Figure 3b. The absolute concentrations of S1 centers (isolated $Nd^{3+}$ centers) show a weak dependence on [$Nd_2O_3$], staying consistently at the level $2–4\times10^{17}/cm^3$ even though the total $Nd_2O_3$ concentration changes by 100×, from $4\times10^{17}/cm^3$ (0.001 mol%) to $4\times10^{19}/cm^3$ (0.1 mol%). The fractional concentration of S1 drops sharply from ~100% in the glass Nd0.001 to 0.3% in Nd0.1. Concurrently, the absolute concentration of S2 centers (EPR-active Nd clusters) shows an almost linear increase with the total $Nd_2O_3$ concentration, from an undetectably low level in the glasses Nd0.001 and LaNd0.001 to about $6–8\times10^{17}/cm^3$ in Nd0.1 and LaNd0.1. The fractional concentration of S2 is remarkably constant at 1–4% through the entire $Nd_2O_3$ concentration range in both $La_2O_3$-free and $La_2O_3$-containing glasses. When comparing $La_2O_3$-free and $La_2O_3$-containing glasses, the addition of 5 mol% $La_2O_3$ consistently suppresses the fractional concentration of S1 centers by 2–3 times, while it has only a marginal effect on the fractional concentration of S2 centers.

Notably, in all glasses (except Nd0.001), the fractional concentrations of S1 and S2 centers, as derived from our EPR analysis, do not sum up to 100% to account for the total Nd concentration in glasses (Figure 3b). The latter suggests that a large fraction of Nd centers stay undetectable in the EPR experiments. Hereafter, these EPR-invisible (undetectable) Nd centers are referred to as the S3 centers. From the measured S1 and S2 fractions in Figure 3b, we infer that the fraction of EPR-invisible S3 centers varies significantly with $La_2O_3$ and $Nd_2O_3$ concentrations in glasses. In glasses with $Nd_2O_3$ concentrations higher than 0.01 mol%, a large fraction of Nd centers (more than 95%) belongs to the category of EPR-invisible S3 centers.

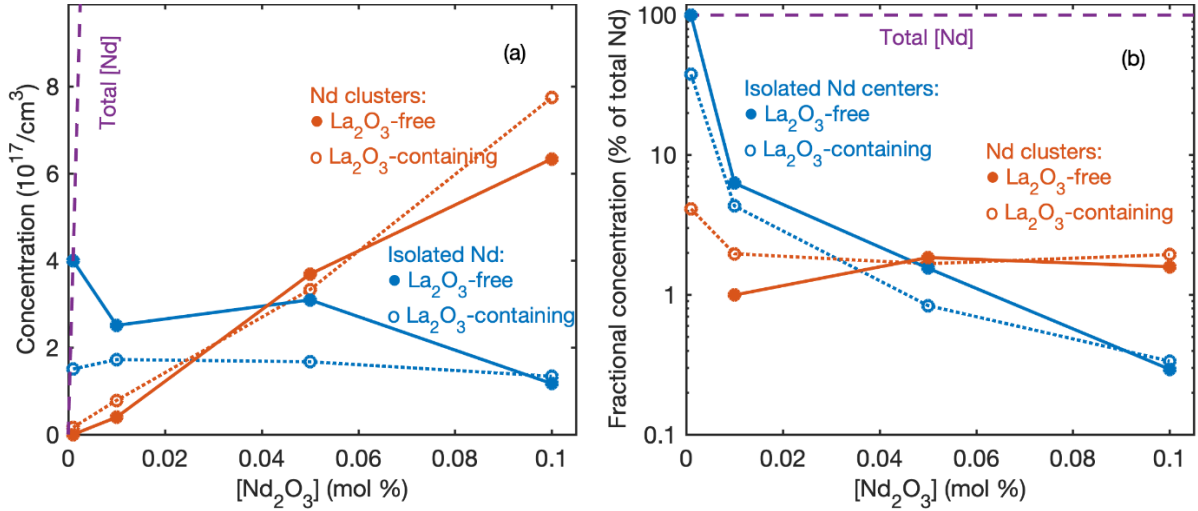

**Figure 3**: Absolute (a) and fractional (b) concentrations of EPR-active Nd centers in $La_2O_3$-free (solid circles, solid lines) and $La_2O_3$-containing (open circles, dashed lines) glasses plotted as a function of the total $[Nd_2O_3]$: (blue) isolated $Nd^{3+}$ centers (the S1 signal), and (red) dipole-coupled Nd clusters (the S2 signal). The vertical, concentration axis in (a) was calibrated by recognizing that all Nd centers in Nd0.001 are isolated and therefore EPR-active, thus contributing 100% to the total EPR signal intensity. The dashed purple lines in (a) and (b) specify the expected total $Nd^{3+}$ concentrations, including both EPR-active and EPR-invisible Nd centers, nominally doped in the glasses.

Based on the above EPR analysis, while quantitatively evaluating both EPR lineshapes and EPR intensities in a series of glasses with varied $Nd_2O_3$ and $La_2O_3$ concentrations (Figure 2 and 3), we, thus, are able to formulate a three-species model for $Nd^{3+}$ centers in peralkaline aluminoborosilicate glasses. The model identifies two EPR-detectable $Nd^{3+}$ centers (S1 and S2), each having a distinct EPR lineshape signature, and also one (or several) EPR-invisible $Nd^{3+}$ species (collectively labeled as S3). The fractional concentrations of all three species in the examined glasses have been determined from our EPR spectral decomposition analysis as a function of $La_2O_3$ and $Nd_2O_3$ concentrations.

*5.1.5   Nature of S2 centers: EPR-visible Nd clusters*

The S2 signal lineshape derived from our EPR decomposition analysis (a yellow trace in Figure 2) differ distinctly from the S1 signal of isolated $Nd^{3+}$ centers (a red trace in Figure 2). The S2 signal has substantially broader linewidth, and its center of gravity is shifted upfield from the S1 signal. As discussed above (in section 5.1.2), similar spectral transformations have been reported in silicate and other glasses for $RE^{3+}$ at elevated concentrations.[3, 5, 34-37, 51, 52] The transformed EPR signals have been usually assigned to either: (1) isolated $RE^{3+}$ centers with coordination environment different from that in glasses with low $RE_2O_3$ concentrations,[34-36, 52] or (2) short-range RE clusters, for example RE–O–RE, where $RE^{3+}$ ions are coupled through strong dipolar interactions.[3, 5, 37, 51] Next, we evaluate these two models as a possible origin of the S2 signal in our glasses.

Examining a possible isolated $Nd^{3+}$ center nature of the S2 signal, we attempted to fit its EPR spectral lineshape assuming an effective spin S = 1/2. The best simulation fit resulted in the anisotropic g-factor values (2.0, 1.0, 0.6). These g-values are unphysical in case of the isolated $Nd^{3+}$ centers and for any possible combinations of ground-state Kramer doublets derived from the $^4I_{9/2}$ multiplet.[49, 50] Such unphysical g-factor values argue strongly against the assignment of the S2 signal to isolated $Nd^{3+}$ centers. Additional argument against the assignment of S2 to isolated $Nd^{3+}$ centers was observed when setting up our pulsed EPR experiments and optimizing the power of microwave pulses to achieve the desired π/2 spin rotations. Given the g-factor values g = (2.0, 1.0, 0.6) for S2 were substantially smaller than g = (3.5, 1.94, 1.0) for S1, the microwave pulses of larger amplitude were expected in case of S2 as compared to S1. Exactly the opposite was observed in our experiments with the optimized pulses being ~1.6× weaker in case of S2, thus arguing against the isolated $Nd^{3+}$ center nature of the S2 signal. On the other hand, this observation of the lower amplitude pulses for S2 was perfectly consistent with the S2 assignment to Nd clusters, as discussed next.

Better EPR fits for the S2 signal lineshape have been obtained using the Nd cluster model. Similar to Sen et al.,[51] the spectral simulations were performed considering on average 4–6 $Nd^{3+}$ ions per cluster, with average distances of 4–5.5 Å between ions within the clusters, while allowing a broad distribution of random asymmetric shapes and topologies of the Nd clusters. Only dipole-dipole interactions between $Nd^{3+}$ spins were allowed within each cluster, and spin-exchange couplings were ignored.[51] These spectral simulations, in combination with the requirement of lower amplitude pulses, leads us to conclude that the S2 signal, as derived from our EPR analysis, should be associated with the dipole-coupled Nd clusters.

The tendency of $RE^{3+}$ to form short-range clusters, with two or more $RE^{3+}$ ions interconnected by oxygens atoms (e.g., RE–O–RE), has been previously reported in pure silica/silicate glasses and other host materials.[1, 3, 51, 53-56] It has been argued that RE clustering in crystals/glasses is thermodynamically favored by the requirement to maintain a local charge balance.[4, 53-56] In alkali aluminoborosilicate glasses, network modifying cations, like $Na^+$, play a major role of charge compensators for trivalent $Al^{3+}$ ($AlO_4^-$) and $B^{3+}$ ($BO_4^-$), each requiring one $Na^+$ for their complete integration into tetrahedral silica-rich network.[19, 20, 38, 57] When $RE^{3+}$ are added to alkali aluminoborosilicate glasses, they also require local charge compensation by $Na^+$ in order to integrate into the glass network.[6, 7] Unlike $Al^{3+}$ and $B^{3+}$, the isolated $RE^{3+}$ ions take more than one $Na^+$, typically 3–4 $Na^+$, for their local charge compensation owing to their preferred octahedral or higher order coordination.[6, 7, 39] In the glasses investigated in the present study, several trivalent cations ($B^{3+}$, $Al^{3+}$, $La^{3+}$, and $Nd^{3+}$) compete for the same limited pool of $Na^+$, and thus a global charge compensation for all cations at once may prove challenging. We, thus, hypothesize that the $RE^{3+}$ clustering could be one of that structural mechanisms through which the global competition for $Na^+$ charge compensation could at least be partially resolved in alkali aluminoborosilicate glasses. Indeed, our ESEEM results presented in section 5.2 (Table 2) show that the Nd-clusters (the S2 centers) require on average fewer numbers of $Na^+$ in their coordination shells as compared to the isolated $Nd^{3+}$ centers (the S1 centers), and thus a large fraction of $Na^+$ is released through the Nd clustering and this $Na^+$ becomes available for charge compensation of other network forming cations.

Three types of short-range Nd clusters can be considered as possible candidates for the S2 signal:[5, 34, 46, 51, 53, 55, 58] (1) the next-neighbor (NN) clusters where two (or more) $Nd^{3+}$ are directly linked to each other through bridging oxygen atoms, e.g., Nd–O–Nd; (2) the next-to-next-neighbor (NNN) clusters where $Nd^{3+}$ are interconnected through a mediator network-forming ion, e.g., Nd–O–X–O–Nd (with X = Si, B, or Al); and (3) the NNN clusters that are interlinked through diamagnetic $La^{3+}$ ions, e.g., Nd–O–La–O–Nd, in case of the $La_2O_3$-doped glasses. From a viewpoint of the EPR experiments, the crucial difference between these three types of Nd clusters is the strength of (anti)ferromagnetic spin-exchange interactions (J) between $Nd^{3+}$ within the clusters. Strong J couplings can have multiple effects on the EPR signal appearance, from slightly broadening the EPR signals to causing their complete disappearance.

The NN clusters as candidates for the S2 signal can be ruled out. Indeed, excessively strong spin-exchange couplings are expected in such short-range clusters, especially for light lanthanide ions like $Nd^{3+}$ and $Ce^{3+}$.[46] Exchange interactions of J = 3–150 GHz have been reported in the NN pairs (Nd–X–Nd, with X = O or F) in several oxide crystals.[46, 53] Interactions of such magnitude would completely broaden out the EPR signals to make them EPR-invisible (see discussion of the EPR-invisible S3 centers in section 5.1.6). The spectral linewidth of the S2 signal (Figure 2) is much narrower than that and is clearly inconsistent with such strong exchange couplings as would be expected in the NN pairs. Therefore, more distant NNN clusters, with substantially weaker spin-exchange interactions, must be considered as candidates for the S2 signal.

As a rule of thumb, the strength of spin-exchange interactions in $RE^{3+}$ pairs scales down by an order of magnitude for each extra bond between $RE^{3+}$.[46, 59] Indeed, much weaker spin-exchange interactions (hundreds MHz) have been reported for $Nd^{3+}$ in distant NNN pairs (Nd–O–X–O–Nd, with X = C, V, or Y), with $Nd^{3+}$ separated by 4.2–7.1 Å.[46, 53, 55] The exchange interactions of this smaller magnitude would be fully consistent with the S2 signal lineshape (Figure 2). We therefore conclude that the long-range NNN clusters, with $Nd^{3+}$ ions interconnected by Nd–O–X–O–Nd linkages, where X is either network forming ($Si^{4+}$, $Al^{3+}$, and $B^{3+}$) or non-framework ($Na^+$) cations, should be considered as the prime candidates for the S2 signal.

The long-range NNN clusters bridged by $La^{3+}$ (Nd–O–La–O–Nd), as may occur in the $La_2O_3$-containing glasses, should be treated separately. They should in fact be assigned to the EPR-undetectable (invisible) S3 centers, and not to the EPR-detectable S2 centers. Although no reports are available on spin exchange couplings of $RE^{3+}$ in the $La^{3+}$-bridged NNN pairs in oxide glasses and crystals, several studies have reported strong (anti)ferromagnetic couplings (J = 3–204 GHz) between organic radicals and transition metal ions bridged via diamagnetic $RE^{3+}$ ions, such as $Y^{3+}$, $La^{3+}$, and $Ho^{3+}$, in organic chain materials.[60-63] These J couplings are comparable in magnitude to J = 3–150 GHz in the short-range NN pairs (Nd–O–Nd).[46, 53] Additional argument in favor of this assignment of the (Nd–O–La–O–Nd) clusters to the EPR-invisible centers is the observation that the number of EPR-detectable S2 centers does not increase (in fact, decreases slightly) in the La$_2$O$_3$-containg glasses as compared to the La$_2$O$_3$-free glasses (Figure 3). In a simple picture, the 50× excess of diamagnetic La$^{3+}$ with respect to paramagnetic Nd$^{3+}$ in the La$_2$O$_3$-containing glasses should be expected to completely dilute a large majority of EPR-invisible NN pairs (Nd–O–Nd) and to convert them into the long-range NNN pairs (Nd–O–La–O–Nd). If these (Nd–O–La–O–Nd) pairs were EPR-detectable S2 clusters, a significant increase in the EPR signal should have been observed which is opposite to what we see in our experiments. This suggests that the 50× dilution of paramagnetic Nd$^{3+}$ with diamagnetic La$^{3+}$ in the La$_2$O$_3$-containing glasses results only in an interconversion of one kind of EPR-invisible S3 centers (Nd–O–Nd) to an another kind of EPR-invisible S3 centers (Nd–O–La–O–Nd), while the number of the EPR-visible S1 centers (isolated Nd$^{3+}$ centers) and S2 centers (Nd–O–X–O–Nd, with X = Si, B, Al) remains low at all times in the glasses.

*5.1.6   Nature of EPR-invisible S3 centers*

The EPR analysis presented in section 5.1.4 reveals that a large fraction of Nd$^{3+}$ in the investigated glasses remains EPR-invisible (referred to as the S3 centers). The fraction of EPR-invisible S3 centers exceeds 90% of total Nd content in the glasses with [Nd$_2$O$_3$] > 0.01 mol%. Here we consider two plausible mechanisms that can turn Nd$^{3+}$ into EPR-invisible entities. The first mechanism is related to a strong spin-exchange interaction between Nd$^{3+}$ within S3 clusters, and the second mechanism refers to the so-called microscopic glass basicity that may favor a partial reduction of trivalent Nd$^{3+}$ into divalent Nd$^{2+}$ in the glasses.[64]

As discussed in section 5.1.5, we identify two types of Nd cluster configurations that can result in strong (anti)ferromagnetic interactions between Nd$^{3+}$ within clusters, thus, making them the EPR-invisible S3 centers. These are: (1) the short-range NN clusters with Nd$^{3+}$ interconnected by bridging oxygens (Nd–O–Nd) in both La$_2$O$_3$-free and La$_2$O$_3$-containing glasses, and (2) the NNN clusters with $Nd^{3+}$ bridged by diamagnetic $La^{3+}$, (Nd–O–La–O–Nd), in the $La_2O_3$-containing glasses. In disordered glassy systems, the clusters can arrange in various shapes and sizes (dimers, trimers, etc.), with a broad distribution of Nd–Nd distances and interconnected topologies, thus, forming a complex network of pairwise spin dipole and exchange interactions. Strong (anti)ferromagnetic exchange interactions within such NN and NNN clusters can make $Nd^{3+}$ spins EPR-invisible in several ways: (1) by pairing up the $Nd^{3+}$ spins in each cluster into EPR-silent singlet states, or more complicated spin-frustrated states, with low lying excited states and therefore, fast longitudinal ($T_1$) spin relaxation; (2) by significantly broadening the EPR spectrum beyond a detectability limit; (3) by introducing additional spin relaxation mechanisms, thus, making longitudinal ($T_1$) or transverse ($T_2$) spin relaxation times of S3 clusters undetectably short. Next, we discuss these mechanisms in some more detail, thus, reaching the conclusion that in order for the S3 centers to be EPR-invisible they have to be the clusters with strong spin-exchange interactions.

Specific to our pulsed EPR experiments, the spin relaxation times of S3 centers would have to be shorter than 50 ns to make them EPR-undetectable. Such short $T_1$ relaxation times are too short to be expected if S3 centers were isolated $Nd^{3+}$ centers or dipole-coupled NNN clusters (Nd–O–X–O–Nd, with X = Si, B, Al). For example, the measured $T_1$ relaxation times for S1 (isolated $Nd^{3+}$ centers) and S2 (dipole-coupled Nd centers) signals resolved in our EPR experiments are both found to be around $T_1 \approx 200$ μs at 4.6 K, with only slight dependence on $Nd_2O_3$ and $La_2O_3$ concentrations in all examined glasses. These long $T_1$ times for S1 and S2 are in full agreement with $T_1 > 100$ μs reported for other $Nd^{3+}$ centers in various host matrices when measured below 5 K.[3, 47, 65, 66] Such long $T_1$ are indeed expected for both isolated $Nd^{3+}$ centers and dipole-coupled Nd clusters since all excited Kramer doublets in $Nd^{3+}$ centers are separated by hundreds of cm$^{-1}$ from the ground state Kramer doublet,[67, 68] and thus all known phonon-assisted $T_1$ spin relaxation processes should be frozen out at temperatures below 5 K.[49] There is only one plausible mechanism that may potentially shorten $T_1$ of S3 centers to below the EPR detectable limit (e.g., $T_1 < 50$ ns, more than 1000× shorter than it is in S1 and S2), and this involves strong spin-exchange couplings between $Nd^{3+}$ spins in S3 clusters. We, therefore, hypothesize that the EPR-undetectable S3 centers in the investigated glasses are short-range NN clusters (Nd–O–Nd) or $La^{3+}$-bridged NNN clusters (Nd–O–La–O–Nd) with strong intra-spin-exchange couplings.

Similar arguments are also valid when examining potential reasons for too short transverse $T_2$ spin relaxation times in S3 centers. The reported values of $T_2$ for the isolated $Nd^{3+}$ centers in different host materials (measured below 5 K) fall in the range between 1 and 100 μs, with shorter $T_2$ observed at higher $Nd_2O_3$ concentrations.[3, 47, 69] Consistently, the $T_2$ measured for S1 (isolated $Nd^{3+}$) and S2 centers (dipole-coupled Nd–O–X–O–Nd clusters) in our glasses are also found to be in the same range, i.e., ~1 μs, with $T_2$ in S2 centers being only slightly shorter (by ~20%) than $T_2$ in S1 (Figure S3). Comparably long $T_2$ on the order of a few μs would also be expected for the S3 centers unless some other $T_2$ relaxation mechanisms are involved, thus shortening their $T_2$ times below the EPR detection limit (e.g., $T_2 < 50$ ns). Strong exchange couplings between $Nd^{3+}$ spins in the NN (Nd–O–Nd) or NNN (Nd–O–La–O–Nd) clusters is one of such mechanisms that can drastically shorten $T_2$, thus making the S3 centers EPR-undetectable.

The second possible mechanism that can lead to the EPR-invisible Nd centers (S3 centers) is the partial reduction of paramagnetic $Nd^{3+}$ to diamagnetic $Nd^{2+}$. In silicate-based glasses, the co-existence of several oxidation states has been demonstrated for a few $RE^{3+}$, such as $Ce^{4+/3+}$, $Sm^{3+/2+}$, $Eu^{3+/2+}$, and $Yb^{3+/2+}$ redox pairs.[70-73] The reduction of $Nd^{3+}$ to $Nd^{2+}$ has also been previously observed in phosphate glasses irradiated with γ rays,[74] but not in silica-based glasses. Based on the literature[6, 7, 27, 75] on $Nd_2O_3$-doped silicate glasses and our own experience,[15] the majority of Nd in the investigated glasses are expected to be in the 3+ state, though the presence of a minor fraction of $Nd^{2+}$ in the glasses cannot be excluded. Therefore, this mechanism seems to be an unlikely cause of generation of S3 centers in the present study.

## 5.2 *Electron Spin Echo Envelope Modulation (ESEEM)*

The ESEEM experiments are performed to probe the nuclear coordination environment of the EPR-active Nd centers (S1 and S2) in $La_2O_3$-free and $La_2O_3$-containing glasses. Figures 4a and Figure 4b present the ESEEM spectra of four glasses (Nd0.001, Nd0.1, LaNd0.001, and LaNd0.1) acquired at two different magnetic fields, 350 and 550 mT. The spectra of glasses containing a low concentration of $Nd_2O_3$ (Nd0.001 and LaNd0.001) measured at both fields 350 and 550 mT represent the S1 centers with only a small admixture (5–10%) from the S2 centers. The situation is more complicated in glasses containing high concentrations of $Nd_2O_3$ (Nd0.1 and LaNd0.1), where the ESEEM spectra are mostly from the S2 centers but with a sizable contribution (15–35%) from the S1 centers, especially when measured at the low field of 350 mT. The extent of contribution from S1 centers can be seen by inspecting the FID-detected spectra and their decompositions into the S1 and S2 states, as presented for Nd0.1 and LaNd0.1 in Figures 2d and 2h, respectively.

The ESEEM spectra resolve several peaks from magnetic isotopes $^{10}B$ (nuclear spin I = 3, natural abundance 20%), $^{11}B$ (I = 3/2, 80%), $^{23}Na$ (I = 3/2, 100%), and $^{27}Al$ (I = 5/2, 100%), as marked in Figure 4. The peaks from $^{23}Na$ and $^{27}Al$ spins are seen to overlap in all the spectra, making their independent analysis difficult. In order to resolve the individual contributions from $^{23}Na$ and $^{27}Al$, we performed an additional ESEEM experiment on a glass with the similar composition as Nd0.01 except that all $Na_2O$ was replaced with an equivalent concentration of $K_2O$ (see Figure A1 and discussion in Appendix A). With $^{23}$Na completely removed from the $K_2O$-containing (KNd0.01) glass we were able to accurately determine the $^{27}$Al coordination in the $Nd^{3+}$ centers, and then these $^{27}$Al parameters were used as fixed while fitting the ESEEM spectra of the Na-containing glasses (like LaNd0.001 and LaNd0.1 in Figure A2).

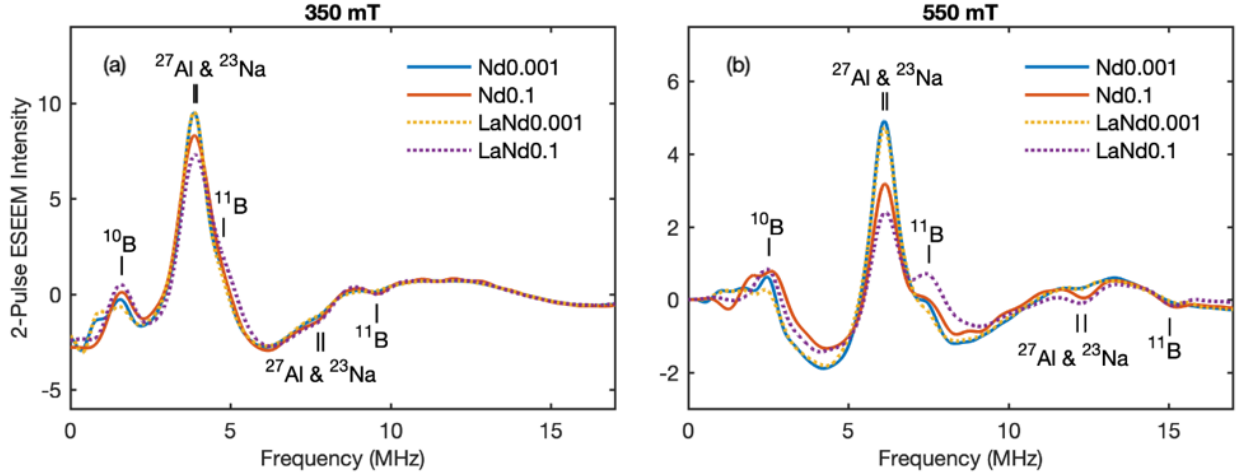

**Figure 4**: Two-pulse FT-ESEEM spectra in $La_2O_3$-free and $La_2O_3$-containing glasses doped with $[Nd_2O_3]$ = 0.001 and 0.1 mol%. The spectra were measured at two magnetic field positions: (a) 350 mT and (b) 550 mT, as identified with the blue circles in Figure 2. The positive (basic) and negative (combination) peaks from $^{10,11}$B, $^{23}$Na and $^{27}$Al nuclear spins are labeled with vertical sticks. All spectra were measured at 4.6 K.

Despite the high concentration of $SiO_2$ in our glasses (> 50 mol%), no peaks from $^{29}$Si could be resolved in the ESEEM spectra (Figure 4). The lack of detectable ESEEM signals from $^{29}$Si can be explained by the low natural abundance (4.7%) of $^{29}$Si isotopes, and by its low nuclear spin (I = ½) as compared to high spins (I = 3/2, 5/2, or 3) for $^{10,11}$B, $^{23}$Na, and $^{27}$Al. Since ESEEM amplitudes generally scale proportionally to I(I+1), much weaker modulation amplitudes are expected in case of $^{29}$Si with its I = ½.[4, 76]

The ESEEM spectra are fitted using the two-shell model developed by Dikanov et al.[76] for situations when hyperfine couplings are weak (see Appendix A for details). Our focus in these simulations is to determine the proximity distances ($R_1$) and the numbers ($N_1$) of magnetic nuclei in the first coordination shell (3–4.5 Å) around $Nd^{3+}$ in the S1 and S2 centers. The ESEEM simulation results are shown in Figures A1-A2 (Appendix), and all parameters derived from the ESEEM fits for the S1 and S2 centers in $La_2O_3$-free and $La_2O_3$-containing glasses are summarized in Table 2. The proximity distances ($R_1$) and the numbers of spins ($N_1$) for $^{10,11}B$ and $^{27}Al$ nuclei are shown as broad ranges, for example, $R_1$ = 3.2–3.6 Å and $N_1$ = 0.4–0.7 in case of $^{27}Al$ nuclei around the S1 centers. These ranges should be understood as the indication that the combinations ($R_1$ = 3.2 Å with $N_1$ = 0.4) and ($R_1$ = 3.6 Å with $N_1$ = 0.7) produce equally good fits to the experimental data. Here, non-integer numbers of $^{27}Al$ spins ($N_1$ = 0.4–0.7) should be understood as: (1) only 40–70% of the S1 centers have one $^{27}Al$ atom in their first coordination shells at a distance $R_1$ = 3.2–3.6 Å, thus forming one Nd–O–Al linkage, and (2) the remaining 30–60% of the S1 centers do not have any $^{27}Al$ atoms in their first coordination shells.

**Table 2**: Proximity distances ($R_1$) and numbers of nuclear spins ($N_1$), including $^{10/11}B$, $^{23}Na$, and $^{27}Al$, in the first (proximal) coordination shell of $Nd^{3+}$ ions in the S1 and S2 centers, as derived from the ESEEM simulations (Figures A1-A2).†

|  |  | $^{10}B$ or $^{11}B$ | $^{23}Na$ | $^{27}Al$ |
|---|---|---|---|---|
| **S1 centers** (Isolated $Nd^{3+}$ centers) | $La_2O_3$-free & $La_2O_3$-containing | $R_1$ = 3.2–3.6 Å $N_1$ = 0.2–0.4 | $R_1$ = 3.2 Å $N_1$ = 4 | $R_1$ = 3.2–3.6 Å $N_1$ = 0.4–0.7 |
| **S2 centers** (Nd clusters) | $La_2O_3$-free | $R_1$ = 3.2–3.6 Å $N_1$ = 0.5–1 | $R_1$ = 3.2 Å $N_1$ = 3 | $R_1$ = 3.2–3.6 Å $N_1$ = 0.5–0.9 |
|  | $La_2O_3$-containing | $R_1$ = 3.2–3.6 Å $N_1$ = 1–2 | $R_1$ = 3.2 Å $N_1$ = 2 | $R_1$ = 3.2–3.6 Å $N_1$ = 0.5–0.9 |

† The two-shell model approximation[76] was used in the ESEEM simulations. In addition to nuclear spins in the first coordination shell around $Nd^{3+}$, the contribution to ESEEM from the second-shell nuclei (remote nuclei > 4.5 Å) was also considered using ($R_2$ = 4.5 Å, $N_2$ = 1.5) for $^{10/11}B$ spins, ($R_2$ = 4.5 Å, $N_2$ = 3.7) for $^{23}Na$ spins, and ($R_2$ = 4.5 Å, $N_2$ = 1.5) for $^{27}Al$ spins. The respective $N_2$ numbers for each isotope were calculated based on the glass composition (Table 1). The same second-shell coordination parameters ($R_2$, $N_2$) were assumed for both S1 and S2 centers.

The ESEEM data (Table 2) reveals that the S1 (isolated $Nd^{3+}$ centers) and S2 centers (dipole-coupled Nd–O–X–O–Nd clusters) have noticeably different first coordination shells. The environment of isolated $Nd^{3+}$ centers (S1 centers) is rich in Na and depleted of B and Al, with on average four Na atoms and fewer than one B/Al atoms in their first coordination shell. Addition of 5 mol% $La_2O_3$ to the glass produces no effect on the environment of S1 centers. As compared to the S1 centers, the environment of the S2 centers (dipole-coupled Nd clusters) is found to be noticeably depleted of Na and enriched in B/Al, and the environment changes further after addition of 5 mol% $La_2O_3$. In $La_2O_3$-free glasses, the S2 centers have on average three Na, one B and one Al, which changes to two Na, two B and one Al in the $La_2O_3$-containing glasses (Table 2).

The distances ($R_1$ = 3.2–3.6 Å) and the numbers ($N_1$ = 0.4–0.9) derived from our ESEEM analysis for $^{27}Al$ coordination around $Nd^{3+}$ centers compare closely with the values reported in the previous ESEEM studies on the $Nd^{3+}$ and $Er^{3+}$-doped silica glasses that were additionally co-doped with small concentrations (1–2 mol%) of $Al_2O_3$.[4, 33] A direct comparison of our results with these previous reports is complicated because of significant differences in our and their glass compositions (e.g., smaller $Al_2O_3$ concentrations, with no $B_2O_3$/$Na_2O$ added, in the previous reports). Furthermore, the $RE^{3+}$ speciation and the fractional concentrations of isolated $RE^{3+}$ centers (S1) versus dipole coupled RE–RE clusters (S2) have not been specified in the previous ESEEM studies. Therefore, their reported coordination numbers ($R_1$, $N_1$) should be viewed as a statistical average over unknown concentrations of isolated $RE^{3+}$ centers and $RE^{3+}$ clusters. As evident from our ESEEM analysis (Table 2), the $RE^{3+}$ coordination environment (the numbers of Al, B and Na atoms in the first coordination shells) varies significantly depending on the RE speciation. Our EPR/ESEEM results call for a re-examination of the previous results on RE-doped silica glasses, including the reported effect of $Al^{3+}$ and $La^{3+}$ co-doping on $RE^{3+}$ clustering,[33, 77] to bring it on a more quantitative ground by accurately determining $RE^{3+}$ speciation in those glasses, as demonstrated in the present work.

As noted above, no ESEEM peaks from $^{29}$Si spins are resolved in our spectra (Figures 4 and A1-A2), and as such no direct spectroscopic information could be obtained about numbers and proximity distances of Si atoms around $Nd^{3+}$ centers. Nevertheless, by assuming a six-fold coordination for $RE^{3+}$ ($Nd^{3+}$, $La^{3+}$) in the glasses, and using the ESEEM-derived coordination numbers of B and Al (Table 2), we anticipate at least 4–5 Si atoms in the first coordination shells of each $Nd^{3+}$ ion in both isolated centers (S1) and dipole-coupled Nd-clusters (S2).

Lastly, we note that our ESEEM spectra in the $La_2O_3$-containing glasses (Figure 4) show no resolvable contribution from $^{139}$La spins, even though the natural abundance of this isotope (99.9%) and its spin (I = 7/2) are both high. The lack of detectable ESEEM signals from $^{139}$La spins in our glasses is consistent with the previous ESEEM reports in the Yb/La-codoped silica glasses where it has been speculated that $La^{3+}$ enters only into the remote 4$^{th}$ coordination shells around paramagnetic $Yb^{3+}$.[77, 78] Alternatively, the lack of detectable $^{139}$La signals may reflect excessively large nuclear quadrupolar (NQI) couplings of $^{139}$La spins. The large NQI would broaden out the ESEEM peaks to make them undetectable, possibly hidden beneath the stronger nearby peaks from $^{10}$B, $^{23}$Na, and $^{27}$Al.

### 5.3 *$^{11}$B and $^{27}$Al Magic Angle Spinning Nuclear Magnetic Resonance (MAS NMR)*

To help interpret the EPR and ESEEM data, we also perform the $^{27}$Al and $^{11}$B MAS NMR experiments on two baseline glasses, i.e., Nd0 ($RE_2O_3$-free) and LaNd0 (containing 5 mol% $La_2O_3$). $La^{3+}$ and $Nd^{3+}$ are often viewed as equivalent in terms of their effect on glass chemistry and network structure.[39, 40, 79] Indeed, both ions have a comparable field strength (FS), e.g., FS($Nd^{3+}$) = 0.54 at $d_{Nd-O}$ = 2.35 Å, versus FS($La^{3+}$) = 0.51 at $d_{La-O}$ = 2.43 Å, and both ions preferably coordinate with 6–8 oxygens in oxide glasses.[39, 40] We recognize that unlike EPR, $^{27}$Al and $^{11}$B NMR experiments probe only the average glass structure from prospects of bulk speciation of network-forming $^{27}$Al and $^{11}$B sites, and as such, these experiments do not provide any direct information about immediate coordination environment around RE$^{3+}$ in glasses. Therefore, our aim in conducting the NMR experiments is to understand the impact of RE$^{3+}$ (5 mol% La$_2$O$_3$) on the structural speciation of network-forming Al and B sites in the glasses and to correlate these NMR-derived structural changes with the RE$^{3+}$ speciation and their coordination environment as determined from the EPR/ESEEM experiments.

### 5.3.1    *$^{27}$Al MAS NMR*

The $^{27}$Al MAS NMR spectra in both Nd0 (La-free) and LaNd0 (5 mol% of La$_2$O$_3$) glasses show a single broad peak centered at around 60 ppm (Figure 5a). Based on the previous NMR studies,[15, 38] the peak is assigned to Al$^{3+}$ ions in a tetrahedral coordination environment (AlO$_4$ sites). Notably, the addition of 5 mol% La$_2$O$_3$ induces no change in the position and linewidth of the AlO$_4$ peak. This observation agrees with other NMR studies on peralkaline aluminoborosilicate glasses with large concentrations of Na$^+$ available to charge compensate all AlO$_4^-$, [Na$_2$O]/[Al$_2$O$_3$] > 1.[6, 7, 40] In such glasses, a single AlO$_4$ peak has been always reported that was insensitive to changes in glass composition, including addition of sizeable concentrations (up to 10 mol%) of non-framework cations, like divalent (Ca$^{2+}$) or trivalent (RE$^{3+}$) ions. This single, composition-insensitive AlO$_4$ peak was interpreted as an evidence of:[6, 7, 38, 40, 80, 81] (1) the negatively charged AlO$_4^-$ sites having a strong preference for charge compensation by weak-field strength cations like Na$^+$ (when available), and only as a second choice by other high-field strength cations, for example, Ca$^{2+}$ or RE$^{3+}$; and (2) the charge-compensated AlO$_4^-$-Na$^+$ units being uniformly dispersed in SiO$_2$ network preferentially linked to four SiO$_4$ units. Taken together these two points explain the insensitivity of the AlO$_4$ peak to changes in glass composition and to incorporation of high field strength cations (RE$^{3+}$) in peralkaline aluminoborosilicate glasses. Specific to our glasses with [Na$_2$O]/[Al$_2$O$_3$] = 2.5, this interpretation implies that an absolute majority of AlO$_4$ in both Nd0 (La$_2$O$_3$-free) and LaNd0 (5 mol% La$_2$O$_3$) glasses maintain the uniform coordination environment being charge compensated by Na$^+$, and only a minor fraction of AlO$_4$ (potentially around 1–2%, if counting the fractions of EPR-detectable S1 and S2 centers as revealed from our EPR experiments) forms the direct Al–O–Nd linkages and thus find itself in a different environment. Such small quantities may be undetectable in $^{27}$Al NMR experiments.

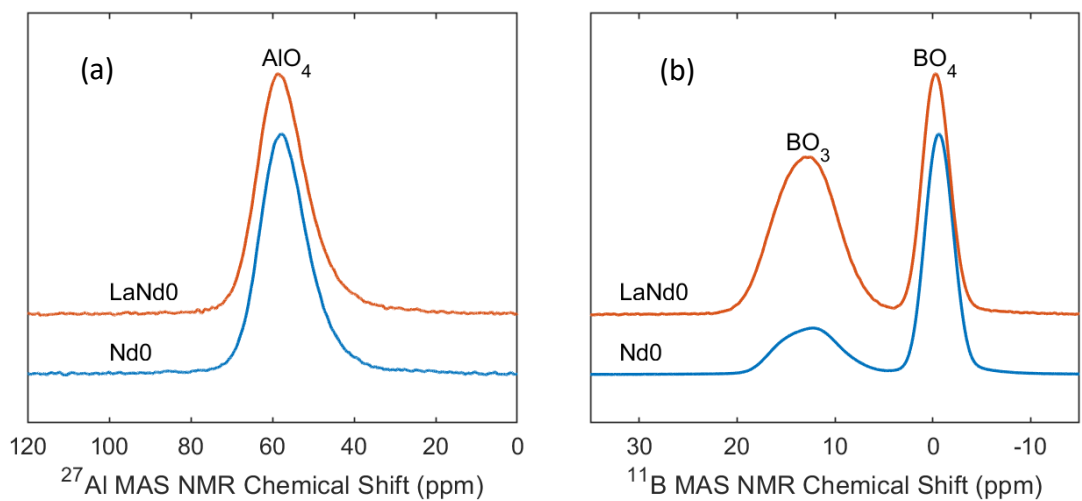

**Figure 5**: $^{27}$Al MAS-NMR (a) and $^{11}$B MAS-NMR (b) spectra of Nd0 (La$_2$O$_3$-free) and LaNd0 (La$_2$O$_3$-containing) glasses measured at 14.1 T. The spectra were normalized to the maximum peak intensity and vertically shifted for clarity. The peak from tetragonal AlO$_4$ sites in (a) and trigonal BO$_3$ and tetragonal BO$_4$ sites in (b) are labeled.

On the other hand, the coordination environment of EPR-invisible S3 centers, that constitute the majority of Nd$^{3+}$ and La$^{3+}$ in our glasses, remains unknown. We hypothesize that the S3 centers might maintain preferentially Al-depleted (possibly Si/B-rich) first coordination shells, and thus the number of direct Al–O–RE linkages remains always low (around 1–2%) even at elevated RE$_2$O$_3$ concentrations, that would explain the insensitivity of the $^{27}$Al NMR signal to addition of 5 mol% La$_2$O$_3$ in LaNd0 (Figure 5a). More systematic and combined NMR/EPR studies would be required to understand the number and nature of direct Al–O–RE linkages in peralkaline alumino-borosilicate glasses.

*5.3.2   $^{11}$B MAS NMR*

The $^{11}$B MAS NMR spectra in glasses Nd0 and LaNd0 resolve two peaks (Figure 5b). The narrow peak at 0 ppm is from tetrahedral borate sites (BO$_4$), and the broad peak at 14 ppm is from several species of trigonal borate sites (BO$_3$).[15, 38, 39] Relative fractions of the BO$_3$ and BO$_4$ sites in the glasses, as determined from the NMR peak integration, are summarized in Table 1. The addition of 5 mol% La$_2$O$_3$ induces a decrease in the fraction of tetrahedral boron (N$_4$) from N$_4$ = 0.66 in Nd0 to 0.37 in LaNd0. Similar interconversion of BO$_4$ to BO$_3$ has been reported in borosilicate and alumino-borosilicate glasses in response to a reduced Na$^+$ concentration or when Na$^+$ was substituted with high-field strength cations like Ca$^{2+}$ or La$^{3+}$.[7, 38-40, 82] In our glasses (Nd0 and LaNd0), the molar ratio $\frac{[Na_2O]}{[Al_2O_3]+[B_2O_3]}$=1.25 is kept constant and high, with more than enough Na$^+$ available to charge compensate all AlO$_4^-$ and BO$_4^-$ sites in both glasses. It is then counterintuitive as to why a simple addition of 5 mol% La$_2$O$_3$ (which potentially is an additional charge compensator for BO$_4^-$ sites) results in a 2× drop of N$_4$ in LaNd0 compared to Nd0?

The answer to this question is provided from our ESEEM experiments (Table 2), showing that Nd$^{3+}$ (and likewise La$^{3+}$) have a strong affinity towards Na$^+$ on their own. Depending on the speciation, each Nd$^{3+}$ ion can immobilize as many as four Na$^+$ if it stays in the isolated S1 form, or two–three Na$^+$ if clustered as the S2 species. Similar conclusions of 3–4 Na$^+$ per RE$^{3+}$ have been reached in the previous NMR studies when interpreting the speciation of network former ions ($^{29}$Si, $^{11}$B, and $^{17}$O) in RE-doped aluminoborosilicate and borosilicate glasses.[7, 27, 39] The NMR analysis was based on simple charge compensation arguments, while additionally postulating that all RE$^{3+}$ ions were isolated (no RE clustering) in the glasses that were doped with high RE$_2$O$_3$ concentrations (2–7 mol%). Even though the assumption of no RE clustering was not well justified, the estimated number of 3–4 $Na^+$ per $RE^{3+}$ was remarkably similar to that directly measured for isolated S1 centers in our ESEEM experiments.

Recognizing the competition between $BO_4^-$, $AlO_4^-$, and $LaO_6^{3-}$ (assuming a six-coordinated $La^{3+}$) for $Na^+$, we can reformulate the concentration of $Na^+$ that remains available for charge compensation of $BO_4^-$ sites in $La_2O_3$-containing glasses as: $[Na_2O]_{ex} = [Na_2O] - [Al_2O_3] - n \cdot [La_2O_3]$. This new definition of $[Na_2O]_{ex}$ is a direct extension of an earlier definition of $[Na_2O]_{ex}$ originated from the NMR studies on sodium borosilicate glasses,[23, 27, 40] now including an additional term that accounts for the effect of $RE^{3+}$ on an overall $Na^+$ availability for $BO_4^-$. The new definition $[Na_2O]_{ex}$ assumes that charge compensation by $Na^+$ follows the order $AlO_4^- > LaO_6^{3-} > BO_4^-$, with $AlO_4^-$ being the strongest competitor for $Na^+$ and $BO_4^-$ being the weakest. Thus, firstly each $AlO_4^-$ site consumes one $Na^+$, followed by each $LaO_6^{3-}$ consuming $n \cdot Na^+$, and finally, the leftover $Na^+$ is used for charge compensation of $BO_4^-$ sites. According to our ESEEM results (Table 2) and past literature,[6, 7, 27, 39] the number (n) of $Na^+$ attracted by each $RE^{3+}$ ion can vary between 2–4 depending on the $RE^{3+}$ speciation in glasses (e.g., isolated $RE^{3+}$ centers vs. RE clusters). Next, we demonstrate the potential of this new definition of $[Na_2O]_{ex}$ in correctly predicting the $N_4$ fraction in our LaNd0 glass.

The ratio $R = [Na_2O]/[B_2O_3]$ has been long identified as one of the key parameters determining the structural organization in ternary borosilicate glasses and also extended to multicomponent glasses like alkali aluminoborosilicates.[39, 82-85] It has been demonstrated that the correctly estimated value of R allows an accurate prediction of fractional populations of $BO_3$ and $BO_4$ units in glasses. Using our definition for $[Na_2O]_{ex}$, we can estimate the ratio $R' = [Na_2O]_{ex}/[B_2O_3] = 1.5$ for Nd0, and $R' = [Na_2O]_{ex}/[B_2O_3] = 0.43$ for LaNd0, simply based on their glass compositions (Table 1). In the case of LaNd0, we assume similar coordination of $La^{3+}$ and $Nd^{3+}$ ions and use the value of n = 2, as derived from our ESEEM experiments for Nd clusters (the S2 centers) in $La_2O_3$-containing glasses (Table 2). Indeed, our EPR experiments (Figure 3b) show that the Nd/La clusters, including the S2 and S3 centers, are the predominant Nd/La species in all LaNdx glasses. Using these estimated R' values and referring to the plot of $N_4$ versus R in Figure S4 derived from the previous NMR studies of ternary alkali borosilicate glasses with high $SiO_2$ content, we can immediately predict the value of $N_4$ = 0.62 for Nd0, and $N_4$ = 0.4 for LaNd0. Both predicted values are in excellent agreement with the $N_4$ numbers measured in our NMR experiments (Table 1), thus demonstrating the predictive accuracy of the new expression for $[Na_2O]_{ex}$. It seems worthwhile examining the predictive utility of this new expression against a broader set of RE-doped aluminoborosilicate glasses.

As discussed in section 5.4.3, the main hurdle in universalizing this new criterion $[Na_2O]_{ex}$ to glasses with varied compositions is the EPR-undetectable S3 clusters with their yet-uncharacterized speciation and coordination environments. Further studies, combining EPR, NMR and other spectroscopic techniques, would be required to generalize and to improve on the accuracy and universality of the new $[Na_2O]_{ex}$ criterion governing the boron speciation in glasses.

As a final note, we would like to emphasize that according to the previous studies, the R' = 0.43 predicted for our LaNd0 glass falls below the critical limit (R = 0.5) where the nano-scale phase separation to Si-rich and B-rich regions may start to develop in sodium borosilicate/aluminoborosilicate glasses depending on their thermal history.[23, 24, 85-87] Indeed, the heat flow versus temperature curves obtained from a differential scanning calorimeter for our LaNd0 glass demonstrated a well-defined crystallization peak ($T_p$ = 1043 K) when being heated above $T_g$ (840 K), while no such peak was observed in the Nd0 glass with R' = 1.5 (Figure S1).

We emphasize that it is the simple addition of 5 mol% $La_2O_3$ and the requirement of minimum two $Na^+$ ions per $RE^{3+}$ that brought the LaNd0 glass to this point of potential immiscibility.

### 5.4  Structural Models for $Nd^{3+}$ centers in sodium aluminoborosilicate glasses

By combining the EPR, ESEEM and NMR results, we propose the following structural models for isolated (S1) and clustered (S2, S3) $Nd^{3+}$ centers in the Ndx and LaNdx glasses.

#### 5.4.1  Isolated $Nd^{3+}$ centers (S1 centers)

As shown by the ESEEM experiments (Table 2), the isolated $Nd^{3+}$ centers (the S1 centers) have similar coordination environments in the $La_2O_3$-free and $La_2O_3$-containing glasses. Thus, each isolated $Nd^{3+}$ center has on average either one $Al^{3+}$ or one $B^{3+}$ in its first coordination sphere along with four $Na^+$ ions. The conclusion of either one $Al^{3+}$ or one $B^{3+}$ is based on the observation that the coordination numbers, $N_1 = 0.7$ for $Al^{3+}$ and $N_1 = 0.4$ for $B^{3+}$, as determined from ESEEM experiments, add up to approximately one for the S1 centers. The potential structure is visualized in Figure 6a, where $Nd^{3+}$ is shown in octahedral coordination, forming six Nd–O–X linkages with the network forming cations X = (Si, Al, or B). As presented, the structure involves five Si linkages and one Al linkage. Alternatively, the Al linkage can be replaced by the B linkage with the relative probability of 0.7:0.4, where the linked $B^{3+}$ or $Al^{3+}$ are in tetrahedral coordination, as is evidenced by our NMR experiments. Along with four $Na^+$ in the coordination sphere of each isolated $Nd^{3+}$ center, the overall net charge of the S1 center is zero. Traditionally, the role of $RE^{3+}$ in silicate glasses is viewed as a network modifier or charge compensator.[6, 7, 12, 24, 39, 88, 89] Interestingly, the structure of the S1 center shown in Figure 6a can also be recognized as a network former where $Nd^{3+}$ is fully polymerized into a glass network as a loosely defined octahedron that is interconnected by oxygen bridges to other (conventional) network formers.

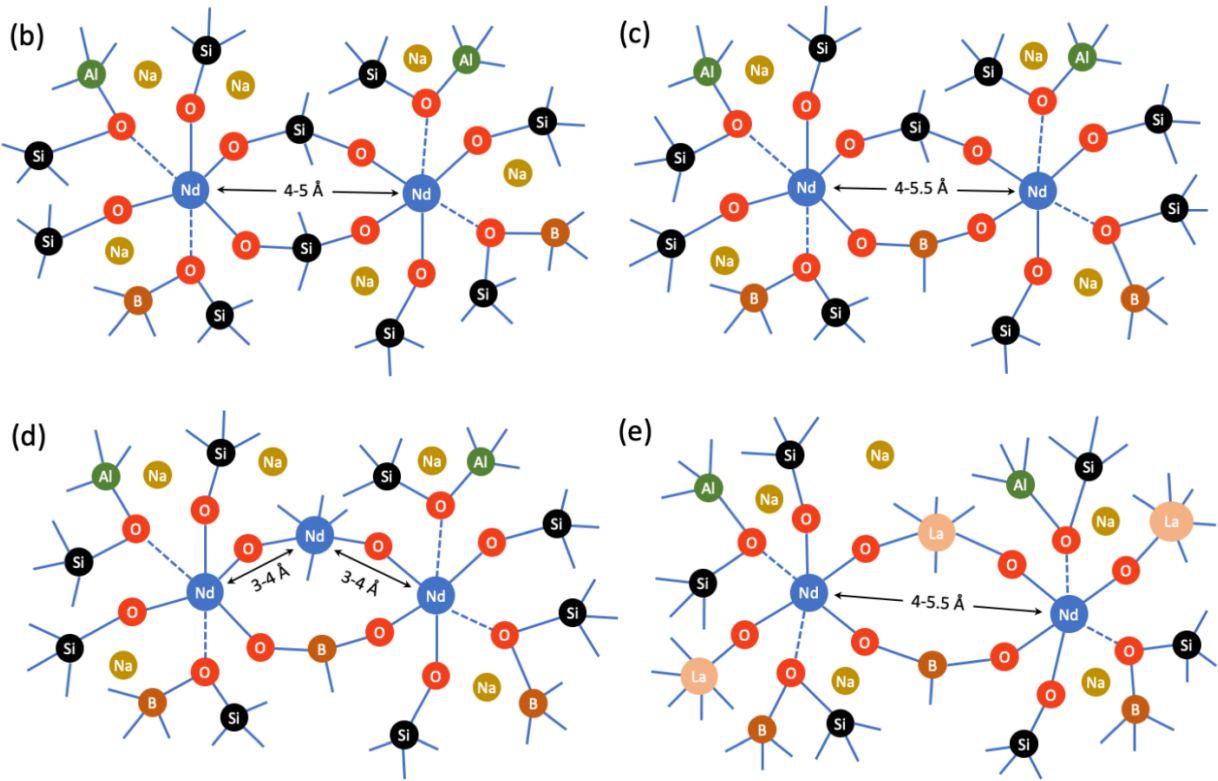

**Figure 6**: Proposed coordination structures for: (a) isolated $Nd^{3+}$ centers (the S1 centers); (b, c) dipole-coupled NNN clusters (the S2 centers) in $La_2O_3$-free (b) and $La_2O_3$-containing (c) glasses, derived from the EPR/ESEEM analysis; (d, e) exchange-coupled, EPR-invisible Nd clusters (the S3 centers) of the NN type (Nd–O–Nd) in $La_2O_3$-free and $La_2O_3$-containing glasses (d), and the NNN type (Nd–O–La–O–Nd) in $La_2O_3$-containing glasses (e), based on discussion in sections 5.1.6 and 5.4.3. Six-fold coordination is assumed for isolated $Nd^{3+}$ centers as well as for each $Nd^{3+}$ ion in Nd clusters, via bridging oxygens (solid lines), or via tri-coordinated oxygens associated with aluminum and boron (dashed lines). For the EPR-active Nd centers (the S1 and S2 centers), the first coordination shell of each $Nd^{3+}$ ion is specified with the numbers of B, Na, and Al atoms as determined from the ESEEM experiments (Table 2). For the EPR-silent Nd clusters (the S3 centers), the first coordination shell environment is extrapolated from the S2 structures thus maintaining the same numbers of B, Na, and Al. The calculated local net charge is zero for the isolated $Nd^{3+}$ centers (a) and for the NNN clusters in $La_2O_3$-free glasses (b), the net charge is -0.5

(per $Nd^{3+}$ ion) in the NNN clusters in $La_2O_3$-containing glasses (c), and it is -1 per $Nd^{3+}$ in both NN and NNN types of the EPR-silent S3 clusters (d, e). Although the structures of the NN and NNN clusters (b-e) are shown as dimers and trimers, they can be easily expanded to incorporate more coupled $Nd^{3+}$ ions within each cluster.

Considering the relative abundance of the network forming cations $Al^{3+}/B^{3+}/Si^{4+}$ = 20/20/55 in the investigated glasses, and under the assumption of their uniform spatial distribution and no preferential coordination at $Nd^{3+}$ sites, we expect on average $N_1$ = 1.3 of the Nd–O–Al linkages and the same number of Nd–O–B linkages around each $Nd^{3+}$. These statistical numbers are noticeably larger than $N_1$ = 0.7 for $^{27}Al$ and $N_1$ = 0.4 for $^{10,11}B$ derived from our ESEEM experiments. We can, thus, conclude that the isolated S1 centers are preferentially formed in a Na/Si-enriched and Al/B-depleted environment in both Ndx and LaNdx glasses. This conclusion seems in conflict with one of the popular viewpoints discussed in the literature,[23-26] where isolated $RE^{3+}$ have been argued to preferentially enter into borate-rich regions, forming RE-metaborate-like structures ($1BO_4$:RE:$2BO_3$) that would imply near 8× higher B coordination numbers from what we measure in our ESEEM experiments.[23, 24, 68]

Our ESEEM results also seem to be in conflict with the earlier reports suggesting that co-doping with near stoichiometric concentrations of $Al^{3+}$ helps improving the $RE^{3+}$ solubility in silica glasses and in particular alleviate the undesired RE clustering.[1] Although the microscopic mechanism of $Al^{3+}$ co-doping on $RE^{3+}$ speciation is not well understood, it has been argued, primarily based on the ESEEM studies, that trivalent $Al^{3+}$ may form the RE–O–Al linkages, presumably substituting some $RE^{3+}$ with $Al^{3+}$ in the RE clusters and effectively decoupling the $RE^{3+}$ ions from each other through what can be thought of as a simple dilution effect.[4] We note however that the earlier studies were performed on the Na-free silicate glasses.[3-5] With a large surplus of $Na^+$ like in glasses examined in our work, it might be thermodynamically advantageous to form isolated $Nd^{3+}$ centers of a kind shown in Figure 6a, with a minimum (or none) Al/B linkage and with four dedicated $Na^+$ ions for charge compensation.

*5.4.2 EPR-active Nd clusters (S2 centers)*

The proposed NNN structures of the EPR-active Nd clusters (the S2 centers) in $La_2O_3$-free and $La_2O_3$-containing glasses are presented in Figure 6b and Figure 6c, respectively. The structures are shown as Nd–Nd pairs, but they can easily be expanded to larger Nd clusters. The $Nd^{3+}$ within the clusters are separated by 4–5.5 Å, bridged by several Nd–O–X–O–Nd linkages (where X = Si, Al, or B, but not $La^{3+}$). These ions are coupled through a weak inter-spin-exchange interaction (therefore, they remain EPR visible) and through a strong spin dipolar interaction (therefore, their broad EPR lineshapes).

According to the ESEEM data (Table 2), the first coordination shell of Nd clusters (the S2 centers) changes noticeably after adding 5 mol% $La_2O_3$. While in $La_2O_3$-free glasses, each $Nd^{3+}$ in the S2 clusters is coordinated by one B, one Al, and three Na ions (Figure 6b), the coordination environment evolves to two B, one Al, and two Na atoms per $Nd^{3+}$ in $La_2O_3$-containing glasses, thus, becoming more enriched in boron and depleted of sodium (Figure 6c). In order to keep the overall net charge low, we assume that the side Nd–O–Al and Nd–O–B linkages involve tri-coordinated oxygens in both $La_2O_3$-free and $La_2O_3$-containing glasses. In a way, each $Nd^{3+}$ in the clusters can be viewed as a charge compensator for one linked $AlO_4^-$ and one $BO_4^-$. At the same time, boron in the bridging position (Nd–O–B–O–Nd) is shown to be in trigonal configuration ($BO_3$) in the $La_2O_3$-containing glasses (Figure 6c). For the structures shown in Figure 6b and 6c, the resulting net charge per $Nd^{3+}$ is 0 and -0.5, respectively. Similar NNN structures have been reported in the RE-containing crystals including $Na_8Y_3(SiO_3)_{12}$ (ID: mp-1202365), $NdAl_3(BO_3)_4$ (ID: mp-6535), and $NaNdB_6O_{13}$ (ID: mp-1198569).[90]

By comparing the S1 structure in Figure 6a with the S2 structure in Figure 6b and then S2 in Figure 6c, we identify two clear tendencies: (1) the Na coordination number per $Nd^{3+}$ gradually goes down, and synchronously (2) the B coordination number grows up. The first tendency is explained by the limited surplus of $Na^+$ in the glasses and by the global competition between several network former cations ($B^{3+}$, $Al^{3+}$ and $RE^{3+}$) for charge compensation by $Na^+$, as discussed in section 5.1.5. Large numbers of $Na^+$ (four $Na^+$ per $RE^{3+}$) are required to charge compensate the isolated S1 centers and to completely integrate them into the glass network, as shown in Figure 6a. As $RE^{3+}$ concentration increases, the problem builds up of insufficient $Na^+$ surplus to fully charge compensate all $RE^{3+}$ and to maintain them in the isolated S1 form. We hypothesize that this $Na^+$ deficiency problem is resolved through RE clustering. Within the RE clusters (Figures 6b, 6c), each $RE^{3+}$ immobilizes only 2–3 $Na^+$ (instead of 4), thus adapting to the $Na^+$ deficiency. We further hypothesize that, instead of $Na^+$, the net charge compensation within the RE clusters is achieved through the enrichment in B coordination numbers. Tri-coordinated oxygens shared with tetrahedral $BO_4^-$ sites and the bridging tri-coordinated boron ($BO_3$) linkages, as shown in Figures 6(b, c), help to maintain the net charge low within the RE clusters. This charge compensation mechanism, utilizing tri-coordinated oxygens, is similar to what is found in peraluminous alkali aluminosilicate glasses ($[Na_2O] < [Al_2O_3]$), where the deficiency in $Na^+$ for $AlO_4^-$ charge compensation is similarly resolved by sharing oxygens between three network forming sites.[91]

As the last note, we emphasize that our ESEEM-derived structures in Figures 6(a-c) are in striking contrast with the popular viewpoint that isolated RE ions enter the B-rich environment, forming RE-metaborate-like structures ($1BO_4$:RE:$2BO_3$),[23-26] and that the RE clusters are preferentially formed in the B-depleted, Si-enriched environment.[23-25, 68] Our ESEEM results show convincingly that the opposite arrangement is actually correct: (1) The isolated $Nd^{3+}$ centers form preferentially in B-depleted, Si-rich glass environment, and (2) the S2 (and possibly S3) clusters form in B-enriched, Si-depleted glass phase. The RE-metaborate-like structures ($1BO_4$:RE:$2BO_3$), as originally proposed for the isolated $RE^{3+}$ centers,[23-26] are in fact more consistent with the S2 clusters.

*5.4.3 Hypothesis-driven structure of EPR-silent Nd clusters (S3 centers)*

Our EPR results reveal that the EPR-silent Nd clusters (the S3 centers) constitute the majority of $Nd^{3+}$ species in both $La_2O_3$-free and $La_2O_3$-containing glasses. The fractional weight of these clusters (S3) grows monotonously with $Nd_2O_3$ concentration, exceeding 95% of total $Nd^{3+}$ in glasses with [$Nd_2O_3$] > 0.05 mol%. Addition of 5 mol% $La_2O_3$ promotes further conversion of a significant fraction (~30–50%) of the isolated $Nd^{3+}$ (S1) centers to the EPR-silent Nd (S3) clusters (Figure 3b), while leaving the fraction of EPR-active clusters (S2) approximately unaffected.

Our EPR experiments provide limited information on the nature of these EPR-silent Nd clusters (S3 centers), only revealing them as EPR-invisible. In our discussion in sections 5.1.5 and 5.1.6, we have identified the main mechanism behind their EPR invisibility, which is the presence of strong (anti)ferromagnetic exchange interactions between $Nd^{3+}$ within the S3 clusters. Two types of exchange-coupled cluster configurations are thus formulated: (1) the short-range NN clusters with $Nd^{3+}$ ions bridged by oxygen, (Nd–O–Nd), in both $La_2O_3$-free and $La_2O_3$-containing glasses, and (2) the long-range NNN clusters with $Nd^{3+}$ ions bridged by diamagnetic $La^{3+}$, (Nd–O–La–O–Nd), in the $La_2O_3$-containing glasses. The respective structures are shown in Figures 6d and 6e. These structures are partially based on the published crystalline structures of $Na_xRE_{(10-x)}Si_6O_{26}$ and $RE_5BSi_2O_{13}$, whose phases are known to nucleate out in RE-doped alkali-borosilicate glasses when exceeding RE solubility limits or after prolonged heat treatments.[23, 24,] [92], [93] Indeed, one of our highly doped glasses (LaNd0.1) was observed to develop a $La_5BSiO_{13}$ crystalline phase after 72 hours of heat-treatment at 865 K (Figure S5).

The proposed NN structure (Nd–O–Nd) in Figure 6d and the La-bridged NNN structure (Nd–O–La–O–Nd) in Figure 6e are both derived from the structures of EPR-active S2 clusters (Figure 6b, 6c) after replacing one of the bridging Nd–O–X–O–Nd linkages with the Nd–O–Nd–O–Nd and Nd–O–La–O–Nd linkages, respectively. In the short-range (Nd–O–Nd) clusters in Figure 6d, the $Nd^{3+}$ are separated by 3–4 Å and linked via direct oxygen bridges, thus, resulting in strong spin-exchange couplings. On the other hand, in the long-range (Nd–O–La–O–Nd) clusters, the strong super-exchange interactions between two remote $Nd^{3+}$ ions propagate via diamagnetic $La^{3+}$ bridges, as hypothesized in section 5.1.5. The La-bridged NNN structure in Figure 6d can also be viewed as part of a larger NN-type cluster where mixed $RE^{3+}$ ions are interconnected through the short-range RE–O–RE linkages. This large cluster consists mostly of diamagnetic $La^{3+}$ but with a small inclusion of paramagnetic $Nd^{3+}$ (e.g., reflecting the 50:1 ratio of La:Nd in our $La_2O_3$-containting glasses). Therefore, direct linkages (Nd–O–Nd) are rare, and most of the linkages in such mixed NN clusters are between mixed RE pairs (Nd–O–La).

Although yet to be determined, the coordination shells of the S3 clusters in Figures 6(d, e) are shown to be enriched in B and partially depleted of Na, as projected from the structurally similar S2 clusters in Figures 6(b, c). The partial Na-depletion of the S3 clusters (with two $Na^+$ per $RE^{3+}$) is consistent with our $^{11}B$ NMR results demonstrating that the $BO_4$ vs $BO_3$ speciation is controlled by a global competition for charge compensation by $Na^+$, through the newly derived expression for $[Na_2O]_{ex}$ with n = 2, as discussed in sections 5.3.2 and 5.4.2. The level of B-enrichment is yet to be confirmed, for example through systematic $^{11}B$ NMR studies.

Since the RE clusters (S2 and S3 in Figures 6) represent a vast majority of RE species (> 90%) in our glasses, their compositional chemistry may play a crucial role in tuning microscopical and mechanical properties of the host glass. The RE clusters formed in Na/B-enriched environment can be viewed as nano-scale-size glass phases, nucleating from the glass and that are enriched in RE, B and Na as compared to the rest of the host glass. The RE/B/Na enrichment in these cluster phases also means the depletion of the same elements from the rest of the host glass. The depletion effect can be small at small RE concentrations ($[Re_2O_3] < 0.1$ mol%), but it can become substantial at higher RE concentrations (like 5 mol% in our $La_2O_3$-containing glasses). Each $RE^{3+}$ within the S2/S3 clusters immobilizes two B and two Na, acting like sponges pulling Na and B from the glass matrix leaving behind a B/Na-depleted glass phase. Therefore, the process of RE cluster formation can be thought as a mechanism inducing a nano-scale phase separation in the glass where RE-rich phase (S2 and S3 clusters) is B/Na-enriched, and the second (RE-free) phase is B/Na-depleted with respect to the nominal chemical composition of the glass. This phase separation as induced by RE clustering can be anticipated to alter mechanical, thermal, and chemical properties of the host glass.

## *6. Summary and Conclusion*

Our combined EPR/NMR investigation of $Nd_2O_3$ and $La_2O_3$ containing peralkaline $Na_2O$-$Al_2O_3$-$B_2O_3$-$SiO_2$ glasses provides several new insights into the partitioning and structural speciation of $RE^{3+}$ in aluminoborosilicate glasses. Our quantitative FID-detected EPR analysis resolves three major Nd species in the glasses: (S1) the isolated $Nd^{3+}$ centers, (S2) the dipole-coupled NNN clusters (Nd–O–X–O–Nd), with $Nd^{3+}$ separated by 4–5.5 Å and interconnected by network-former bridges (X = Si, B, Al), and (S3) the exchange-coupled, EPR-invisible Nd clusters that are further subdivided into two categories of the NN clusters (Nd–O–Nd) separated by 3–4 Å and the La-bridged NNN clusters (Nd–O–La–O–Nd) separated by 4–5.5 Å. The fractional populations of these three RE species as extracted from the EPR spectral decompositions are found to depend strongly on total RE concentrations in the glass. The isolated centers (S1) are only abundant at low [RE$_2$O$_3$] < 0.01 mol%, and the NN and NNN clusters (S2 and S3) prevail at higher concentrations, exceeding 99% of total RE in glasses with [RE$_2$O$_3$] ≥ 0.1 mol%. For example, at [Nd$_2$O$_3$] = 0.1 mol% the average distance of 30 Å is expected between Nd ions assuming their uniform (no clustering) distribution in the glass. The Nd–Nd distances of 3–5.5 Å observed in the NN and NNN clusters are much shorter than that, thus emphasizing highly non-uniform distribution of RE ions in the glasses and their high tendency to form clusters with much shorter RE–RE distances.

The ESEEM experiments provide further insight into the near shell coordination environment of the EPR-detectable Nd$^{3+}$ species (S1 and S2) in the investigated glasses. Thus, on average four Na and one Al (or one B) are found in the 1$^{st}$ coordination shells of isolated Nd$^{3+}$ centers (S1), and the numbers change to fewer Na (two–three), more B (one–two) and more Al (one) in the coordination shells of each Nd$^{3+}$ in the NNN clusters (S2). The significant observation here is a substantial change in the preferred coordination environment from isolated Nd$^{3+}$ centers (S1) to the NNN clusters (S2), with fewer Na and more B involved in case of the clusters. This observation provides us an important clue on the possible mechanism behind the high tendency of RE$^{3+}$ to cluster in the glasses. Indeed, in order to form one isolated RE$^{3+}$ center (S1) requires a large number (four) of Na$^+$ ions for its charge compensation. Recognizing an additional (strong) competition for the same Na$^+$ pool from network-forming AlO$_4^-$ and BO$_4^-$, one may expect that four Na$^+$ are not easily available for each RE$^{3+}$ (especially at high RE$_2$O$_3$ concentrations, like 5 mol% in our La$_2$O$_3$-containing glasses), and therefore the isolated Nd$^{3+}$ centers (S1) cannot be formed. We hypothesize that the charge compensation problem for RE$^{3+}$ ions in glass matrix is resolved through their clustering. Indeed, our ESEEM data confirm that as little as two $Na^+$ per each $RE^{3+}$ are sufficient for their charge compensation. We further hypothesize that instead of $Na^+$ the charge compensation in RE clusters is achieved through the B enrichment of their 1st coordination shells (as evident from the ESEEM data) and by employing the tri-coordinated oxygen structures with $AlO_4$ and $BO_4$ (Figure 6b-e), similar to reported in per-aluminous silicate glasses with $[Na_2O] < [Al_2O_3]$ [91], along with $BO_3$–bridged RE–RE linkages.

The tendency of $RE^{3+}$ to cluster in B/Na-enriched environment has an important consequence on the overall glass structure. In a way, the $RE^{3+}$ clusters can be viewed as initial seeds of a nano-scale glass phase separation, where one (cluster) phase is enriched in $RE^{3+}$, $B^{3+}$ and $Na^+$, and the second (leftover) phase is depleted of the same components. This RE-induced nano-scale phase separation can be anticipated to have a significant effect on physico-chemical and mechanical properties of host glasses.

*Appendix A. Details of ESEEM simulations for LaNd0.001, LaNd0.1 and K-Nd0.01.*

The ESEEM simulations are performed using the two-shell model developed by Dikanov et. al.[76] Accordingly, all nuclear spins around the S1 and S2 centers are subdivided into two subgroups. One group comprises of the proximal nuclear spins at distances of 3–4.5 Å, that define the first and second coordination shells around the Nd centers, and the second group accounts for more distant nuclear spins (> 4.5 Å). The numbers of spins ($N_2$) in the second group (the distant shell) are assumed to be fixed during the simulations, and they are calculated as $N_2 = (4\pi/3)\rho R_2^3$, using the average density ($\rho$) of the respective nuclei in the glass and by setting the proximity distance $R_2$ at 4.5 Å.[76] Of primary interest in our two-shell model simulations are the nuclei in the first (proximal) shell. Thus, the proximity distance ($R_1$) and the numbers of spins ($N_1$) in the proximal shell are two fitting parameters during the ESEEM simulations. We note that these two parameters ($R_1$, $N_1$) may show some interdependence since the integrated ESEEM amplitude generally scales as a product of $N_1/R_1^6$.

All nuclear spins contributing to the ESEEM effects in our glasses, including $^{10,11}$B, $^{23}$Na, and $^{27}$Al, have their spins I > ½. Therefore, their nuclear quadrupolar interactions (NQI) should be accounted during the simulation. The best fits to our experimental ESEEM data are obtained using the NQI parameters: ($C_Q \equiv e^2Qq/h$ = 4 MHz, $\eta$ = 0) for $^{27}$Al, (4 MHz, $\eta$ = 0.5) for $^{23}$Na, and (1.5 MHz, $\eta$ = 0.1) for $^{11}$B. The NQI parameters derived for $^{27}$Al and $^{23}$Na are in close agreement with the values reported in the NMR experiments for alkali aluminoborosilicate glasses.[38, 40, 94] However, the NQI parameter $C_Q$ = 1.5 MHz determined for $^{11}$B is noticeably smaller than $C_Q$ = 2.6–2.7 MHz typically reported from the NMR experiments for trigonal $BO_3$ sites, and larger than $C_Q$ < 0.5 MHz for tetrahedral $BO_4$ sites.[40, 82] We speculate that the difference in our ESEEM-derived $C_Q$ may reflect a strong effect on $^{11}$B quadrupolar coupling from $Nd^{3+}$ coordination when forming the Nd–O–B linkages. Lastly, in our ESEEM simulations, for all involved nuclear spins we assume a model of random relative orientations of NQI tensors with respect to the respective hyperfine coupling tensors as appropriate for disordered glass systems.

To help resolving the overlapping signals from $^{23}$Na and $^{27}$Al spins, and thus to allow their independent evaluation, we also perform an additional ESEEM experiment on a glass with the same composition as Nd0.01 but with all $Na_2O$ being replaced with an equivalent concentration of $K_2O$ (the respective glass is labeled as K-Nd0.01). Figure A1 shows the ESEEM spectra and the fits for K-Nd0.001. The first coordination shell parameters ($R_1$, $N_1$) for $^{27}$Al nuclei in the S1 centers are thus independently determined from the ESEEM fit of K-Nd0.01, where no interference from $^{23}$Na is present. These $^{27}$Al parameters are then used as fixed parameters when fitting the ESEEM effects in the sodium-containing glasses (like LaNd0.001 and LaNd0.1 in Figure A2), where both $^{27}$Al and $^{23}$Na contribute to the spectra.

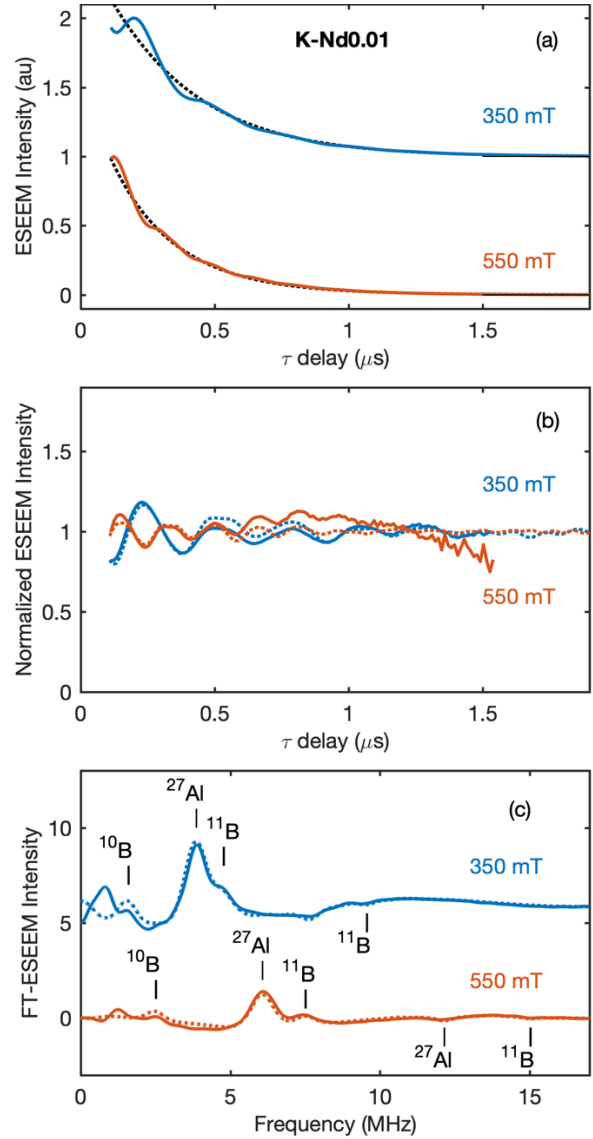

**Figure A1**: Two-pulse ESEEM for K-Nd0.01, where 25 mol% of Na$_2$O were replaced with equivalent amount of K$_2$O. The dominant Nd$^{3+}$ centers in K-Nd0.01 were isolated S1 centers. (a) The ESEEM decays (solid lines) measured at two field positions, 350 and 550 mT, and their exponential fits (dashed lines). (b) The normalized ESEEM time-domains (solid lines), obtained by dividing the measured decays from (a) with their exponential fits. The dashed lines are the ESEEM simulations assuming that each isolated Nd$^{3+}$ center (the S1 center) has on average 0.7 of $^{27}$Al spins at a distance 3.6 Å, and 0.4 of $^{10}$B/$^{11}$B spins at 3.6 Å (Table 2). (c) Phase-corrected Fourier Transform (FT) spectra of the experimental (solid lines) and simulated (dashed lines) ESEEM time-domains from (b) showing the peaks from $^{27}$Al, $^{10}$B, and $^{11}$B spins.

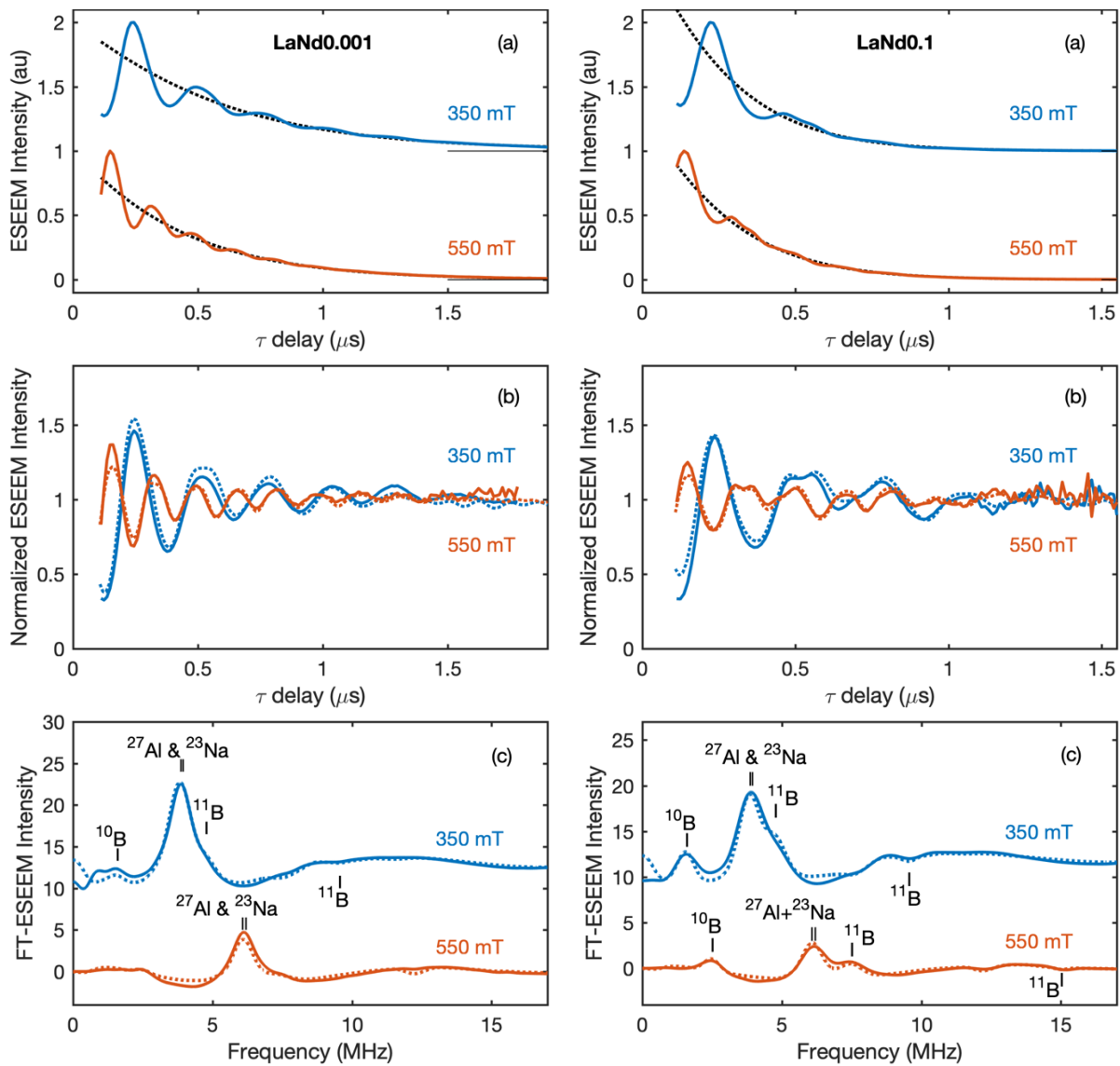

**Figure A2**: Two-pulse ESEEM for LaNd0.001 (left panels) and LaNd0.1 (right panels), where the dominant Nd species were isolated $Nd^{3+}$ centers (the S1 centers) and dipole-coupled Nd clusters (the S2 centers), respectively. (a) The ESEEM decays (solid lines) measured at two field positions, 350 and 550 mT, marked with blue circles on the EPR spectra in Figures 2(e, h). The dashed black curves are exponential fits to the measured decays. (b) The normalized ESEEM time-domains (solid lines), obtained after dividing the measured decays (a) by their exponential fits. The dashed lines are the ESEEM simulations assuming that: (left) each isolated $Nd^{3+}$ center (the S1 center) has on average 0.7 of $^{27}Al$ spins at a distance 3.6 Å, four $^{23}Na$ spins at 3.2 Å, and 0.4 of $^{10}B/^{11}B$ spins at 3.6 Å (Table 2), and (right) each $Nd^{3+}$ ion in dipole-coupled Nd clusters (the S2 centers) has on average 0.9 of $^{27}Al$ spins at a distance 3.6 Å, two $^{23}Na$ spins at 3.2 Å, and two $^{10}B/^{11}B$ spins at 3.6 Å. (c) Phase-corrected Fourier Transformation (FT) spectra of the experimental (solid lines) and simulated (dashed lines) ESEEM time-domains from (b) showing strong overlapping peaks from $^{23}Na$ and $^{27}Al$ spins and smaller peaks from $^{10}B$ and $^{11}B$ as labeled.

**Acknowledgements**

This research is being performed using funding received from the U.S. Department of Energy – Office of Nuclear Energy through the Nuclear Energy University Program under the award DE-NE0008431 and DE-NE0008597. F. Wang acknowledges the support from Southwest University of Science and Technology, China, and China Scholarship Council for funding his visit to Rutgers University.

**Supplementary Information**

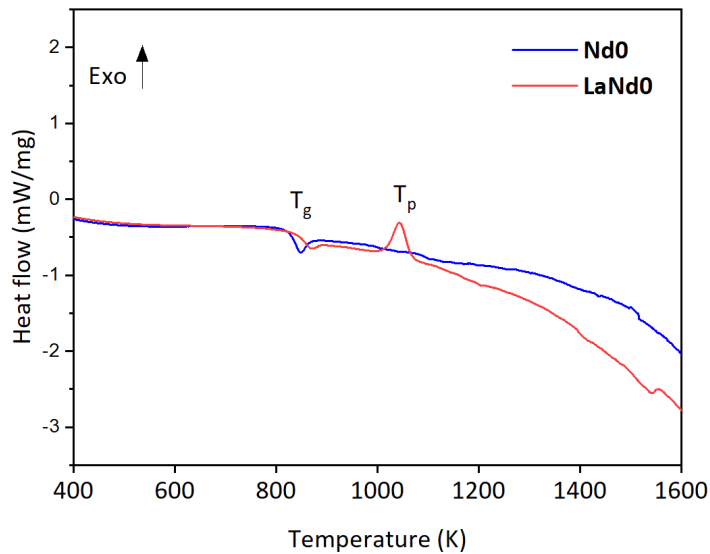

**Figure S1.** Heat flow versus temperature curves for **Nd0** ($La_2O_3$-free) and **LaNd0** ($La_2O_3$-containing) glasses obtained from a differential scanning calorimeter (DSC). The exothermic phase transitions are observed by a positive shift (Exo ↑). $T_p$ refers to the peak temperature of crystallization (an exothermic process) and is only observed in the LaNd0 glass ($T_p$ = 1043 K). Both glasses exhibit the characteristic glass transition endotherm (negative dip), the onset of which is reported as the glass transition temperature ($T_g$). $T_g$ = 824 K for Nd0 and 840 K for LaNd0.

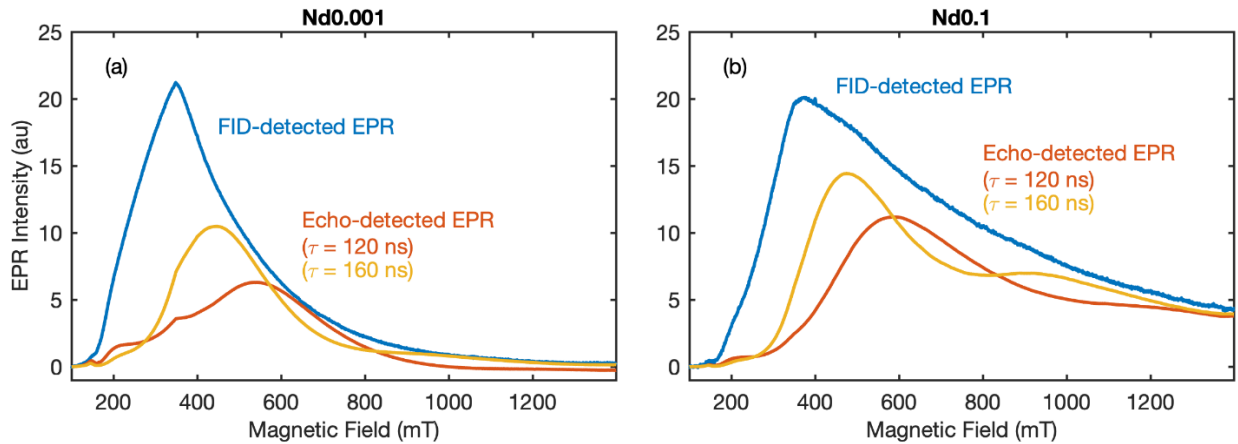

**Figure S2.** Comparison of the FID-detected and the echo-detected EPR spectra measured for two glasses, **Nd0.001** (a) and **Nd0.1** (b). The echo-detected EPR spectra were measured using two-pulse spin echo experiment with inter-pulse delays set to τ = 120 and 160 ns (red and yellow traces, respectively), by integrating the entire echo signal while sweeping through magnetic field. All spectra were measured at 4.6 K. When compared to the FID-detected EPR (blue traces), the echo-detected EPR (red and yellow traces) demonstrates highly distorted spectral lineshapes and a significant loss in signal intensities. These spectral distortions originate from the interfering, field-dependent ESEEM effects due to hyperfine interactions with magnetic nuclei, like $^{10,11}$B, $^{23}$Na and $^{27}$Al, in the glasses.[1-3] The extent of spectral distortion depends on choice of the inter-pulse delay τ. On the other hand, the FID-detected EPR spectra (blue traces) are completely free from any ESEEM-related spectral distortions. The highly accurate EPR lineshapes and intensities measured with the FID-detected EPR provided a ground for the quantitative EPR analysis of $Nd^{3+}$ speciation in the RE-doped glasses investigated in this work.

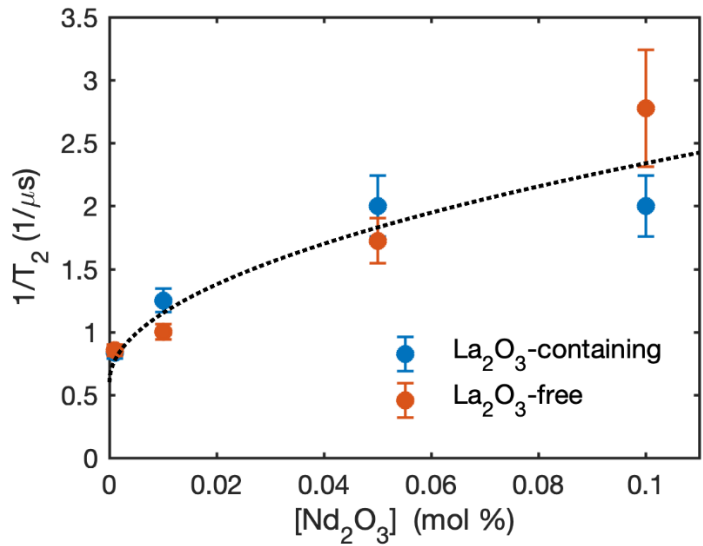

**Figure S3.** Nd/La concentration dependence of spin decoherence times ($T_2$) for EPR-active $Nd^{3+}$ centers in La-free and La-containing glasses measured at 4.6 K. The $T_2$ times were measured at magnetic field 350 mT where the EPR signal was mostly from the S1 centers (isolated $Nd^{3+}$ centers) at low $[Nd_2O_3] < 0.05$ mol%, and the signal had a sizable contribution from the S2 centers (Nd clusters) at high $[Nd_2O_3] > 0.05$ mol% (cp. Figure 2). The dashed curve is a fit to the data using the equation $1/T_2 = C + R \cdot \sqrt{[Nd_2O_3]}$, that takes into account two spectral diffusion mechanisms: the concentration-independent term ($C$) is from flip-flops of environmental nuclear spins,[4, 5] and the square-root dependent term ($R$) is from neighboring $Nd^{3+}$ spins undergoing slow spin relaxation.[6]

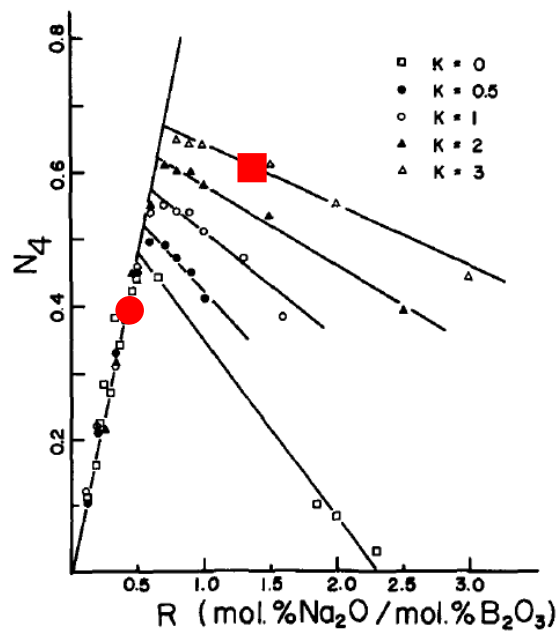

**Figure S4.** Variation of a tetrahedral boron fraction (N$_4$) as a function of R = [Na$_2$O]/[B$_2$O$_3$] in sodium borosilicate glasses. *Reprinted from Yun, Y. H.; Bray, P. J*[7] *with permission from Elsevier*. The estimated R' values for two glasses examined in this work are R' = 1.25 for **Nd0**, and R' = 0.43 for **LaNd0**, as calculated using the glass compositions (Table 1) and the new expression for R' introduced by us in the main text, [Na$_2$O]$_{ex}$ = [Na$_2$O] – [Al$_2$O$_3$] – n·[La$_2$O$_3$], with n = 2. The calculated K = [SiO$_2$]/([B$_2$O$_3$] + [Al$_2$O$_3$]) is equal to 2.75 in both glasses. The respective points, Nd0 (■) and LaNd0 (●), are shown on the plot predicting N$_4$ = 0.62 for Nd0 and N$_4$ = 0.4 for LaNd0, in good agreement with the values 0.66 and 0.37, respectively, measured in our NMR experiments.

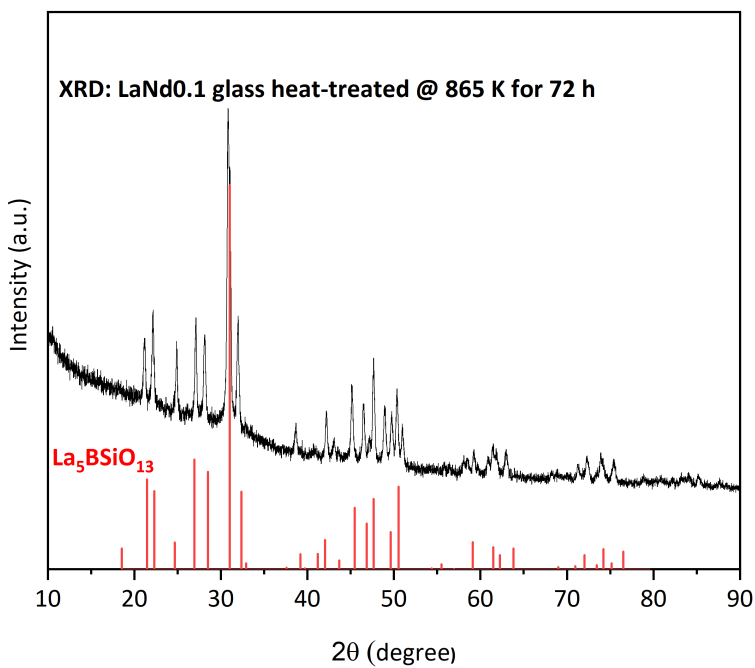

**Figure S5**. X-ray diffraction pattern of glass LaNd0.1 heat-treated at 865 K ($T_g + 50$ K) for 72 hours. The pattern exhibits sharp peaks suggesting that the glass has undergone crystallization. The peaks are identified with powder diffraction file (PDF) database. The identified phase is lanthanum boron silicate ($La_5BSiO_{13}$ – PDF # 00-052-0699).